\documentclass[12pt,a4paper]{article}
\usepackage[top=2.3cm,bottom=2.7cm,left=2.2cm,right=2.2cm]{geometry}
\usepackage{latexsym}
\usepackage{amsmath}
\usepackage{amsfonts}
\usepackage{amssymb}
\usepackage{amsthm}
\usepackage{amscd}
\usepackage{mathrsfs}

\newcommand{\be}{\begin{equation}}
\newcommand{\ee}{\end{equation}}
\newcommand{\bea}{\begin{eqnarray}}
\newcommand{\eea}{\end{eqnarray}}

\renewcommand{\theequation}{\arabic{section}.\arabic{equation}}

\def\1{{\boldsymbol 1}}                %
\def\0{{\boldsymbol 0}}                %
\def\blambda{{\boldsymbol\lambda}}     %
\def\cA{{\mathcal A}}                  %
\def\cC{{\mathcal C}}                  %
\def\cD{{\mathcal D}}                  %
\def\cF{{\mathcal F}}                  %
\def\cG{{\mathcal G}}                  %
\def\cH{{\mathcal H}}                  %
\def\cH{{\mathcal H}}                  %
\def\cO{{\mathcal O}}                  %
\def\cR{{\mathcal R}}                  %
\def\cZ{{\mathcal Z}}                  %
\def\cS{{\mathcal S}}                  %
\def\red{\mathrm{red}}                 %
\def\tr{\mathrm{tr}}                   %
\def\diag{\mathrm{diag}}               %
\def\diag{{\rm diag}}                  %
\def\id{{\rm id}}                      %
\def\red{{\rm red}}                    %
\def\reg{{\rm reg}}                    %
\def\rd{{\rm d}}                       %
\def\ri{{\rm i}}                       %
\def\sgn{{\rm sgn}}                    %
\def\tr{{\rm tr}}                      %
\def\vreg{{\rm sreg}}                  %
\def\Re{{\rm Re}}                      %
\def\C{\mathbb{C}}                     %
\def\N{\mathbb{N}}                     %
\def\R{\mathbb{R}}                     %
\def\T{{\mathbb T}}                    %
\def\GL{{\rm GL}}                      %
\def\UN{{\rm U}}                       %
\def\gl{\mathfrak{gl}}                 %
\def\un{\mathfrak{u}}                  %
\def\fL{\mathfrak{L}}                  %
\def\fQ{\mathfrak{Q}}                  %
\begin{document}

\begin{center}
{\large\bf
Duality between the trigonometric $\boldsymbol{\mathrm{BC}_n}$ Sutherland system
and a completed rational Ruijsenaars\,--\,Schneider\,--\,van Diejen system}
\end{center}

\vspace{0.2cm}

\begin{center}
L.~Feh\'er${}^{a,b}$ and T.F.~G\"orbe${}^a$\\

\bigskip

${}^a$Department of Theoretical Physics, University of Szeged\\
Tisza Lajos krt 84-86, H-6720 Szeged, Hungary\\
e-mail: tfgorbe@physx.u-szeged.hu

\medskip
${}^b$Department of Theoretical Physics, WIGNER RCP, RMKI \\
H-1525 Budapest, P.O.B.~49,  Hungary\\
e-mail: lfeher@physx.u-szeged.hu

\bigskip

\bigskip

\end{center}

\vspace{0.2cm}

\begin{abstract}
We present a new case of duality between integrable many-body systems, where two
systems live on the action-angle phase spaces of each other in such a way that
the action variables of each system serve as the particle positions of the other one.
Our investigation utilizes an idea that was exploited previously to
provide group-theoretic interpretation for several dualities discovered originally
by Ruijsenaars.
In the group-theoretic framework one applies Hamiltonian reduction to two Abelian
Poisson algebras of invariants on a higher dimensional phase space and identifies their reductions
as action and position variables of two integrable systems living on two different models of
the single reduced phase space. Taking the cotangent bundle of $\mathrm{U}(2n)$ as the upstairs space,
we demonstrate how this mechanism leads to a new dual pair involving the $\mathrm{BC}_n$ trigonometric
Sutherland system. Thereby we generalize earlier results
pertaining to the $\mathrm{A}_n$ trigonometric
Sutherland system as well as a recent work
by Pusztai on the hyperbolic $\mathrm{BC}_n$ Sutherland system.
\end{abstract}

\tableofcontents

\newpage

\section{Introduction}
\setcounter{equation}{0}

The integrable one-dimensional many-body systems of Calogero\,--\,Sutherland\,--\,Toda type
and their generalizations are very important because they are ubiquitous in
physical applications and have close ties to several topics of mathematics.
See, for example, the reviews \cite{Eti, Nekr, Per,  Banff, Suth, Toda}.
We here focus on their fascinating duality relations, which were first studied by
Ruijsenaars \cite{SR88}.
We shall uncover a new case of duality
between two systems of this type.

Duality between two Liouville integrable Hamiltonian systems $(M,\omega,H)$ and
$(\tilde{M},\tilde{\omega},\tilde{H})$ requires the existence of Darboux coordinates
$q_i,p_i$ on $M$ and $\lambda_j,\vartheta_j$ on $\tilde M$ (or on dense open
submanifolds of $M$ and $\tilde{M}$) and a \emph{global} symplectomorphism
$\cR\colon M\to\tilde{M}$ such that $(\lambda, \vartheta) \circ \cR$ are action-angle variables
for the Hamiltonian $H$ and $(q, p) \circ \cR^{-1}$ are action-angle variables for the
Hamiltonian $\tilde H$. This means that $H \circ \cR^{-1}$ depends only
on $\lambda$ and $\tilde H \circ \cR$ only on  $q$.
Then one says that $(M, \omega,  H)$ and
$(\tilde M, \tilde \omega, \tilde H)$ are in action-angle duality.
In addition, for the systems of our interest it also happens
that when expressed in the coordinates $(q,p)$ the Hamiltonian $H(q,p)$ admits
interpretation in terms of interaction of $n$ `particles' with position variables $q_i$, and
 $\tilde H(\lambda,\vartheta)$ similarly describes the interaction of $n$ points with positions
$\lambda_i$. Thus the $q_i$ are particle positions for $H$ and action variables
for $\tilde H$, and the $\lambda_i$ are positions for $\tilde H$ and actions for $H$.
The significance of  this curious property is clear for instance from
the fact that it persists at the quantum mechanical level as the
bispectral character of the wave functions \cite{DG,SR-Kup}, which are important special functions.

Dual pairs of many-body systems were exhibited by Ruijsenaars
in the course of his direct construction of action-angle variables for the many-body systems
(of non-elliptic Calogero\,--\,Sutherland type and non-periodic Toda type)  associated with
the $\mathrm{A}_n$ root system \cite{SR88,SR90,RIMS95,Banff}.
It is natural to expect that
action-angle duality exists also for many-body
systems associated with other root systems.  Substantial evidence to support this expectation
was given in a recent paper by Pusztai \cite{P3}, where  action-angle
duality between the hyperbolic $\mathrm{BC}_n$ Sutherland  \cite{IM, OP-Inv}
and the rational Ruijsenaars\,--\,Schneider\,--\,van Diejen (RSvD) systems \cite{vD}
was established.
The specific goal of the present work is to find out how this result can be generalized if one
replaces the hyperbolic $\mathrm{BC}_n$ system with its trigonometric analogue.
A similar problem has been studied previously
in the $\mathrm{A}_n$ case, where it was found that the
dual of the trigonometric Sutherland system possesses intricate
global structure \cite{FA, RIMS95}. The global description of the duality necessitates a separate
investigation also
in the  $\mathrm{BC}_n$ case, since
it cannot be derived by naive analytic continuation between trigonometric and
hyperbolic functions.
This problem turns out to be considerably more complicated than those studied
in \cite{FA,P3}.

The trigonometric $\mathrm{BC}_n$ Sutherland system is defined by the Hamiltonian
\be
H(q,p)=\frac{1}{2}\sum_{j=1}^np_j^2
+\sum_{1\leq j<k\leq n}\bigg[\frac{\gamma}{\sin^2(q_j-q_k)}
+\frac{\gamma}{\sin^2(q_j+q_k)}\bigg]
+\sum_{j=1}^n \frac{\gamma_1}{\sin^2(q_j)}
+\sum_{j=1}^n \frac{\gamma_2}{\sin^2(2q_j)}.
\label{I1}
\ee
Here $(q,p)$ varies in the cotangent bundle $M=T^*C_1=C_1\times\R^n$ of the domain
\be
C_1=\bigg\{q\in\R^n\bigg|\frac{\pi}{2}>q_1>\cdots>q_n>0\bigg\},
\label{I2}
\ee
and the three independent real coupling constants $\gamma,\gamma_1,\gamma_2$
are supposed to satisfy
\be
\gamma>0,\quad
\gamma_2>0,\quad
4\gamma_1+\gamma_2>0.
\label{I3}
\ee
The inequalities in (\ref{I3}) guarantee that the $n$ particles with coordinates $q_j$ cannot
leave the open interval
$(0, \frac{\pi}{2})$ and they cannot collide.
At a `semi-global' level, the dual system will be shown to have the Hamiltonian
\bea
&&\tilde{H}^0(\lambda,\vartheta)=\sum_{j=1}^n\cos(\vartheta_j)
\bigg[1-\frac{\nu^2}{\lambda_j^2}\bigg]^{\tfrac{1}{2}}
\bigg[1-\frac{\kappa^2}{\lambda_j^2}\bigg]^{\tfrac{1}{2}}
\prod_{\substack{k=1\\(k\neq j)}}^n
\bigg[1-\frac{4\mu^2}{(\lambda_j-\lambda_k)^2}\bigg]^{\tfrac{1}{2}}
\bigg[1-\frac{4\mu^2}{(\lambda_j+\lambda_k)^2}\bigg]^{\tfrac{1}{2}}\nonumber\\
&&\qquad \quad \qquad -\frac{\nu\kappa}{4\mu^2}\prod_{j=1}^n
\bigg[1-\frac{4\mu^2}{\lambda_j^2}\bigg]
+\frac{\nu\kappa}{4\mu^2}.
\label{I4}
\eea
Here $\mu>0$, $\nu$, $\kappa$ are real constants,
$\vartheta_1,\ldots,\vartheta_n$ are angular variables,
and $\lambda$ varies in the Weyl chamber with thick walls:
\be
C_2=\bigg\{\lambda\in\R^n\bigg|
\begin{matrix}\lambda_a-\lambda_{a+1}>2\mu,\\
(a=1,\ldots,n-1)\end{matrix}
\quad\text{and}\quad
\lambda_n>\max\{|\nu|,|\kappa|\}\bigg\}.
\label{I5}
\ee
The inequalities defining $C_2$ ensure the reality and the smoothness of $\tilde H^0$ on
the phase space $\tilde M^0:= C_2 \times\T^n$, which is equipped with the symplectic form
\be
\tilde\omega^0=\sum_{k=1}^n\rd\lambda_k\wedge\rd\vartheta_k.
\ee
Duality will be established under the following relation between the coupling parameters,
\be
\gamma=\mu^2,\quad
\gamma_1=\frac{\nu\kappa}{2},\quad
\gamma_2=\frac{(\nu-\kappa)^2}{2},
\label{I6}
\ee
where in addition to $\mu>0$ we also adopt the condition
\be
\nu > \vert \kappa \vert \geq 0.
\label{I7}
\ee
This  entails that
equation (\ref{I6}) gives a one-to-one correspondence between the parameters
$(\gamma, \gamma_1, \gamma_2)$ subject to (\ref{I3}) and $(\mu, \nu, \kappa)$, and
also serves to simplify our analysis.
In the above,  the qualification `semi-global' indicates that
$\tilde M^0$ represents a dense open submanifold
of the full dual phase space, $\tilde M$.
The completion of $\tilde M^0$ into $\tilde M$ guarantees both the completeness of the Hamiltonian
flows of the dual system and the global nature of the
symplectomorphism between $M$ and $\tilde M$.
The structure of $\tilde M$ will be clarified in the paper. For example, we shall see
that the action variables of the  Sutherland system fill
the closure of the domain $C_2\subset\R^n$, with the boundary points corresponding to degenerate
Liouville tori.

The integrable systems $(M,\omega, H)$ and $(\tilde M, \tilde \omega, \tilde H)$
as well as their duality relation will emerge from an appropriate
Hamiltonian reduction.
Specifically, we will reduce
the cotangent bundle $T^*\mathrm{U}(2n)$ with respect to the symmetry group $G_+ \times G_+$, where
$G_+\cong\mathrm{U}(n)\times\mathrm{U}(n)$ is the fix-point subgroup of an involution of $\mathrm{U}(2n)$.
This enlarges the range of the reduction approach to action-angle
dualities \cite{JHEP,Gors,Nekr}, which realizes \cite{Feh-Tod, FA,FK-JPA,FK-CMP,FK-NPB}
the following scenario.
Pick a higher dimensional symplectic  manifold $(P,\Omega)$ equipped with two
Abelian Poisson algebras $\fQ^1$ and $\fQ^2$ formed by invariants under a symmetry group
acting on $P$. Then perform Hamiltonian reduction leading to the reduced manifold
$(P_\red, \Omega_\red)$  carrying the reduced Abelian Poisson algebras $\fQ^1_\red$ and
$\fQ^2_\red$.
Under favorable circumstances, it is possible to construct two models $(M,\omega)$ and
$(\tilde M, \tilde \omega)$ of $(P_\red, \Omega_\red)$
in such a way that when expressed in terms of  $(M,\omega)$ $\fQ^1_\red$ and $\fQ^2_\red$
coincide with the Abelian Poisson algebras generated by the position and action variables
of an integrable many-body Hamiltonian $H$, respectively, and one finds a similar picture
 from the dual perspective of
$(\tilde M, \tilde \omega, \tilde H)$
except that the roles of $\fQ^1_\red$ and $\fQ^2_\red$ are interchanged.
In particular, the many-body Hamiltonian $H$ on $M$ is engendered by an element of
$\fQ^2_\red$ and the many-body Hamiltonian $\tilde H$ on $\tilde M$ is born from
an element of $\fQ^1_\red$.
For a relatively simple and enlightening example, we recommend the reader
to have a glance at
the duality between the hyperbolic
$\mathrm{A}_n$ Sutherland and rational Ruijsenaars\,--\,Schneider  systems as described in
\cite{FK-JPA}.

The rest of the paper is organized as follows.
In the next section, we present the necessary group-theoretic preliminaries together
with the definition of the unreduced Abelian Poisson algebras $\fQ^1, \fQ^2$ and the
symplectic reduction to be performed.
Then Section 3 is devoted to the derivation of
the first model $(M,\omega)$ of the reduced phase space that carries
the Sutherland Hamiltonian obtained as the reduction of the free Hamiltonian
governing geodesic motion on $\mathrm{U}(2n)$.
The content of this section, and even its quantum analogue,  is fairly standard \cite{FP-RMP}.
The heart of the paper is Section 4, where we develop the dual model
$(\tilde{M},\tilde{\omega})$
of the reduced phase space and explain how the Hamiltonian $\tilde{H}$ arises.
This section relies on a blend of ideas from \cite{FA} and \cite{P1,P2,P3}, and also
requires the solution of a number of  rather non-trivial technical problems.
Some technical details are relegated to an appendix.
Our main new results are  given by Theorem 4.1 and Theorem 4.10,
which yield, respectively, the `semi-global' and a fully global characterization
of the reduced phase space.
Finally, in Section 5, we pull together the previous developments and
discuss the duality between the two systems mentioned in the title of the paper.
Here, we shall also use the action angle-duality to establish
interesting properties of these Hamiltonian systems.

\section{Preparations}
\setcounter{equation}{0}

We next describe the starting data which will lead to integrable many-body systems
in duality by means of the mechanism outlined in the Introduction.
We then summarize some group-theoretic facts that will be used in the demonstration of
this claim.

\subsection{Definition of the Hamiltonian reduction}

Let us choose an arbitrary positive integer, $n$,
and also introduce $N:=2n$.
Our investigation requires the unitary group of degree $N$
\be
G:=\UN(N)=\{y\in\GL(N,\C)\mid y^\dag y=\1_N\},
\label{P1}
\ee
and its Lie algebra
\be
\cG:=\un(N)=\{Y\in\gl(N,\C)\mid Y^\dag+Y=\0_N\},
\label{P2}
\ee
where $\1_N$ and $\0_N$ denote the identity and null matrices of size $N$,
respectively. We endow the Lie algebra $\cG$
with the Ad-invariant bilinear form
\be
\langle\cdot,\cdot\rangle\colon\cG\times\cG\to\R,\quad
(Y_1,Y_2)\mapsto\langle Y_1,Y_2\rangle:=\tr(Y_1Y_2),
\label{P3}
\ee
and identify $\cG$ with the dual space
$\cG^\ast$ in the usual manner.
By using left-translations to trivialize the cotangent
bundle $T^*G$, we also adopt the identification
\be
T^\ast G\cong G\times\cG^\ast\cong G\times\cG
=\{(y,Y)\mid y\in G,\ Y\in\cG\}.
\label{P4}
\ee
Then the canonical symplectic form of $T^\ast G$
can be written as
\be
\Omega^{T^*G}:=-\rd\langle y^{-1}\rd y,Y\rangle.
\label{P5}
\ee
It can be evaluated according to the formula
\be
\Omega^{T^*G}_{(y,Y)}(\Delta y\oplus\Delta Y,\Delta'y\oplus\Delta'Y)
=\langle y^{-1}\Delta y,\Delta'Y\rangle
-\langle y^{-1}\Delta'y,\Delta Y\rangle
+\langle[y^{-1}\Delta y,y^{-1}\Delta'y],Y\rangle,
\label{P6}
\ee
where $\Delta y\oplus\Delta Y,\Delta'y\oplus\Delta'Y\in T_{(y,Y)} T^*G$
are arbitrary tangent vectors at a point $(y,Y)\in T^*G$.

Let us introduce the $N\times N$ Hermitian, unitary matrix
partitioned into four $n\times n$ blocks
\be
C:=\begin{bmatrix}
\0_n&\1_n\\
\1_n&\0_n
\end{bmatrix}\in G,
\label{P7}
\ee
and the involutive automorphism of $G$ defined as conjugation with $C$
\be
\Gamma\colon G\to G,\quad
y\mapsto\Gamma(y):=CyC^{-1}.
\label{P8}
\ee
The set of fix-points of $\Gamma$ forms the subgroup of $G$
consisting of $N\times N$ unitary matrices with centro-symmetric block structure,
\be
G_+=\{y\in G\mid\Gamma(y)=y\}=\bigg\{
\begin{bmatrix}
a&b\\
b&a
\end{bmatrix}\in G\bigg\}\cong\UN(n)\times\UN(n).
\label{P9}
\ee
We also introduce the closed submanifold $G_-$ of $G$ by
the definition
\be
G_-=\{y\in G\mid\Gamma(y)=y^{-1}\}=\bigg\{
\begin{bmatrix}
a&b\\
c&a^\dag
\end{bmatrix}\in G\bigg\vert\ b,c\in\ri\un(n)\bigg\}.
\label{P10}
\ee
By slight abuse of notation,  we let $\Gamma$ stand for the induced
involution of the Lie algebra $\cG$, too. We can decompose $\cG$ as
\be
\cG=\cG_+\oplus\cG_-,\quad Y=Y_++Y_-,
\label{P11}
\ee
where $\cG_\pm$ are the eigenspaces of $\Gamma$
corresponding to the eigenvalues $\pm 1$, respectively, i.e.,
\be
\begin{split}
\cG_+&=\ker(\Gamma-\id)=\bigg\{
\begin{bmatrix}
A&B\\
B&A
\end{bmatrix}
\bigg\vert\ A,B\in\un(n)\bigg\},\\
\cG_-&=\ker(\Gamma+\id)=\bigg\{
\begin{bmatrix}
A&B\\
-B&-A
\end{bmatrix}
\bigg\vert\ A\in\un(n),\ B\in\ri\un(n)\bigg\}.
\end{split}
\label{P12}
\ee
We are interested in a reduction of $T^*G$ based on the symmetry group
$G_+ \times G_+$.  We shall use the shifting trick of symplectic reduction \cite{OR},
and thus we first prepare a coadjoint orbit of the symmetry group.
To do this, we take any vector $V\in \C^N$ that satisfies $CV+V=0$, and associate
to it the element $\upsilon_{\mu,\nu}^\ell(V)$ of $\cG_+$ by the definition
\be
\upsilon_{\mu,\nu}^\ell(V):=\ri\mu\big(VV^\dag-\1_N\big)+\ri(\mu-\nu)C,
\label{P13}
\ee
where $\mu,\nu\in\R$ are real parameters.
The set
\be
\cO^\ell :=
\big\{\upsilon^\ell\in\cG_+\mid
\exists\ V\in\C^N,\ V^\dag V=N,\ CV+V=0,\
\upsilon^\ell=\upsilon_{\mu,\nu}^\ell(V)
\big\}
\label{P14}
\ee
represents a coadjoint orbit of $G_+$ of dimension $2(n-1)$.
We let $\cO^r:=\{\upsilon^r\}$ denote the one-point coadjoint orbit of $G_+$
containing the element
\be
\upsilon^r:=-\ri\kappa C\quad\text{with some constant}\;\kappa\in\R,
\label{P15}
\ee
and consider
\be
\cO:=\cO^\ell\oplus\cO^r\subset\cG_+\oplus\cG_+\cong(\cG_+\oplus\cG_+)^\ast,
\label{P16}
\ee
which is a coadjoint orbit\footnote{The same coadjoint orbit was used in \cite{P3}.} of $G_+ \times G_+$.
Our starting point for symplectic reduction will be the phase space
$(P,\Omega)$ with
\be
P :=T^\ast G\times\cO
\quad\mbox{and}\quad
\Omega:=\Omega^{T^*G}+\Omega^{\cO},
\label{P17}
\ee
where $\Omega^{\cO}$ denotes the Kirillov\,--\,Kostant\,--\,Souriau
symplectic form on $\cO$. The natural symplectic action of
$G_+\times G_+$ on $P$ is defined by
\be
\Phi_{(g_L,g_R)}(y,Y,\upsilon^\ell\oplus\upsilon^r)=
\big(g_L^{\phantom{1}}yg_R^{-1},
     g_R^{\phantom{1}}Yg_R^{-1},
     g_L^{\phantom{1}}\upsilon^\ell g_L^{-1}\oplus
     \upsilon^r \big).
\label{P18}
\ee
The corresponding  momentum map $J\colon P\to\cG_+\oplus\cG_+$
is given by the formula
\be
J(y,Y,\upsilon^\ell\oplus\upsilon^r)=
\big((yYy^{-1})_++\upsilon^\ell\big)\oplus\big(-Y_++\upsilon^r\big).
\label{P19}
\ee
We shall see that the reduced phase space
\be
P_\red=P_0/(G_+\times G_+),\qquad
P_0:=J^{-1}(0),
\label{P20}
\ee
is a smooth symplectic manifold, which inherits two
Abelian Poisson algebras from $P$.

Using the identification $\cG^*\cong\cG$, the invariant functions
$C^\infty(\cG)^G$ form the center of the Lie\,--\,Poisson bracket.
Denote by $C^\infty(G)^{G_+\times G_+}$ the set of smooth functions on $G$
that are invariant
under the $(G_+\times G_+)$-action on $G$ that appears in the first component of (\ref{P18}).
Let us also introduce the maps
\be
\pi_1\colon P\to G,
\quad
(y,Y, \upsilon^\ell, \upsilon^r)\mapsto y,
\label{P21}\ee
and
\be
\pi_2\colon P\to\cG,
\quad
(y,Y, \upsilon^\ell, \upsilon^r)\mapsto Y.
\label{P22}
\ee
It is clear that
\be
\mathfrak{Q}^1:= \pi_1^*(C^\infty(G)^{G_+\times G_+})
\quad\hbox{and}\quad
\mathfrak{Q}^2:= \pi_2^*( C^\infty(\cG)^G)
\label{P23}
\ee
are two Abelian subalgebras in the Poisson algebra of smooth functions on $(P,\Omega)$
and these Abelian Poisson algebras descend to the reduced phase space $P_\red$.

Later we shall construct two models of $P_\red$ by exhibiting
two global cross-sections for the action of $G_+ \times G_+$ on $P_0$.
For this, we shall apply two different methods for solving
the constraint equations that, according to \eqref{P19}, define
the level surface $P_0\subset P$:
\be
(yYy^{-1})_++\upsilon^\ell=\0_N
\quad\mbox{and}\quad
-Y_++\upsilon^r=\0_N,
\label{P24}
\ee
where $\upsilon^\ell=\upsilon^\ell_{\mu,\lambda}(V)$ \eqref{P13} for some
vector $V\in\C^N$ subject to $CV+V=0$, $V^\dag V=N$ and $\upsilon^r = - \ri \kappa C$.
We below collect the group-theoretic results needed for our constructions.

\subsection{Recall of group-theoretic results}

To start, let us associate the diagonal $N\times N$ matrix
\be
Q(q):=\diag(q,-q)
\label{P25}
\ee
with any $q\in\R^n$.  Notice that the set
\be
\cA:=\{\ri Q(q)\mid q\in\R^n\}\subset\cG_-
\label{P26}
\ee
is a maximal Abelian subalgebra in $\cG_-$.
The corresponding subgroup of $G$ has the form
\be
\exp(\cA)=\big\{
e^{\ri Q(q)}=
\diag\big(e^{\ri q_1},\ldots,e^{\ri q_n},
          e^{-\ri q_1},\ldots,e^{-\ri q_n}\big)
\mid q\in\R^n\big\}.
\label{P27}
\ee
The centralizer of $\cA$  inside $G_+$ \eqref{P9} (with respect to conjugation)
is the Abelian subgroup
\be
Z:=Z_{G_+}(\cA)=\big\{
e^{\ri \xi}=
\diag\big(e^{\ri x_1},\ldots,e^{\ri x_n},
          e^{\ri x_1},\ldots,e^{\ri x_n}\big)
\mid x\in\R^n\big\}<G_+.
\label{P28}
\ee
The Lie algebra of $Z$ is
\be
\cZ=\{\ri \xi=\ri\, \diag(x,x)\mid x\in\R^n\} <\cG_+.
\label{P29}
\ee

The results that we now recall (see e.g. \cite{Helg,Mat,Sch})  will be used later.
First, for any $y\in G$ there exist elements $y_L$, $y_R$ from $G_+$ and
unique $q\in \R^n$ satisfying
\be
\frac{\pi}{2} \geq q_1 \geq \cdots \geq q_n \geq 0
\label{P30}
\ee
such that
\be
y=y_L^{\phantom{1}}e^{\ri Q(q)}y_R^{-1}.
\label{P31}
\ee
If all components of
$q$ satisfy strict inequalities, then the pair $y_L, y_R$ is unique precisely up to
the replacements $(y_L,y_R)\to(y_L\zeta,y_R\zeta)$ with arbitrary $\zeta\in Z$.
The decomposition (\ref{P31}) is referred to as the generalized Cartan decomposition corresponding to the
involution $\Gamma$.

Second, every element $g \in G_-$ can be written in the form
\be
g = \eta e^{2\ri Q(q)} \eta^{-1}
\label{P32}
\ee
with some $\eta\in G_+$ and uniquely determined $q\in\R^n$ subject to (\ref{P30}).
In the case of strict inequalities for $q$, the freedom in $\eta$ is given precisely by
the replacements $\eta\to\eta\zeta,\ \forall\,\zeta\in Z$.

Third, every element $Y_-\in\cG_-$ can be written in the form
\be
Y_-=g_R\ri D g_R^{-1},\qquad
D=\diag(d_1,\ldots, d_n,-d_1,\ldots, - d_n),
\label{P33}
\ee
with $g_R \in G_+$ and uniquely determined real $d_i$ satisfying
\be
d_1  \geq \cdots \geq d_n \geq 0.
\label{P34}
\ee
If the $d_i$ satisfy strict inequalities, then the  freedom
in $g_R$ is exhausted by the replacements $g_R\to g_R\zeta$, $\forall\,\zeta\in Z$.

The first and the second statements are essentially
equivalent since the map
\be
G \to G_-, \quad y \mapsto y^{-1} C y C
\label{P35}
\ee
descends to a diffeomorhism from
\be
G/G_+ = \{ G_+ g \mid g\in G\}
\label{P36}
\ee
onto $G_-$ \cite{Helg}.

\section{The Sutherland picture}
\setcounter{equation}{0}

We here exhibit a symplectomorphism between the reduced phase space $(P_\red, \Omega_\red)$
and the Sutherland phase space
\be
M=T^* C_1 = C_1 \times \R^n
\label{S1}
\ee
equipped with its canonical symplectic form, where $C_1$ was defined in (\ref{I2}).
As preparation, we associate with any $(q,p)\in M$ the $\cG$-element
\be
Y(q,p):=K(q,p)-\ri\kappa C,
\label{S2}
\ee
 where $K(q,p)$ is the $N\times N$ matrix
\be
\begin{gathered}
K_{j,k}=-K_{n+j,n+k}=\ri p_j\delta_{j,k}-\mu(1-\delta_{j,k})/\sin(q_j-q_k),\\
K_{j,n+k}=-K_{n+j,k}=(\nu/\sin(2q_j)+\kappa\cot(2q_j))\delta_{j,k}
+\mu(1-\delta_{j,k})/\sin(q_j+q_k),\\
\end{gathered}
\label{S3}
\ee
with $j,k=1,\ldots,n$.
We also introduce the $N$-component vector
\be
V_\R:=(\underbrace{1,\ldots,1}_{n\ {\rm times}},
\underbrace{-1,\ldots,-1}_{n\ {\rm times}})^\top.
\label{S4}
\ee
Notice from  (\ref{P12}) that $K(q,p) \in \cG_-$.

Throughout the paper we adopt the conditions (\ref{I7}) and take $\mu>0$, although
the next result requires only that the real parameters
$\mu, \nu, \kappa$ satisfy
\be
\mu \neq 0
\quad\hbox{and}\quad
\vert \nu \vert \neq  \vert \kappa \vert.
\label{S5}
\ee

\noindent \bf\sffamily Theorem 3.1. \it
Using the notations introduced in \eqref{P13}, \eqref{P25} and \eqref{S2},
the subset $S$ of the phase space $P$ \eqref{P17} given by
\be
S:=\left\{ (e^{\ri Q(q)},Y(q,p),\upsilon_{\mu,\nu}^\ell(V_\R),\upsilon^r) \mid
(q,p) \in M \right\},
\label{S6}
\ee
is a global cross-section for the action of $G_+ \times G_+$ on $P_0= J^{-1}(0)$.
Identifying $P_\red$ with $S$, the reduced symplectic form is equal to the Darboux form
$\omega= \sum_{k=1}^n\rd q_k \wedge\rd p_k$.
Thus the obvious identification between $S$ and $M$ provides a symplectomorphism
\be
(P_\red,\Omega_\red)\simeq(M,\omega).
\label{S7}
\ee
\rm
\begin{proof}
We saw in Section 2 that the points of the level surface $P_0$
satisfy the equations
\be
(yYy^{-1})_++\upsilon^\ell_{\mu,\nu}(V)=\0_N
\quad\hbox{and}\quad
-Y_+ -\ri \kappa C=\0_N,
\label{S8}
\ee
for some vector $V\in\C^N$ subject to $CV+V=0$, $V^\dag V=N$.
Remember that the block-form of any Lie algebra element $Y\in\cG$ is
\be
Y=\begin{bmatrix}A&B\\-B^\dag&D\end{bmatrix}
\quad\mbox{with}\quad
A+A^\dag=\0_n=D+D^\dag,\quad
B\in\C^{n\times n}.
\label{S9}
\ee
Now the second constraint equation in
\eqref{S8} can be written as
\be
2Y_+
=\begin{bmatrix}A+D&B-B^\dag\\B-B^\dag&A+D\end{bmatrix}
=\begin{bmatrix}\0_n&-2\ri\kappa\1_n\\-2\ri\kappa\1_n&\0_n\end{bmatrix}
=-2\ri\kappa C,
\label{S10}
\ee
which implies that
\be
D=-A
\quad\mbox{and}\quad
B^\dag=B+2\ri\kappa\1_n.
\label{S11}
\ee
Thus every point of $P_0$ has $\cG$-component $Y$ of the form
\be
Y=\begin{bmatrix}A&B\\-B-2\ri\kappa\1_n&-A\end{bmatrix}
\quad\mbox{with}\quad
A+A^\dag=\0_n,
\quad
B\in\C^{n\times n}.
\label{S12}
\ee
By using the generalized Cartan decomposition (\ref{P31}) and applying a gauge transformation
(the action of $G_+ \times G_+$ on $P_0$), we may assume that $y = e^{\ri Q(q)}$ with
some $q$ satisfying (2.30).
Then the first equation of the momentum map constraint \eqref{S8}
yields the matrix equation
\be
\frac{1}{2\ri}\big(e^{\ri Q(q)}Ye^{-\ri Q(q)}+e^{-\ri Q(q)}CYCe^{\ri Q(q)}\big)
+\mu(VV^\dag-\1_N)+(\mu-\nu)C=\0_N.
\label{S13}
\ee
If we introduce the notation $V=(u,-u)^\top$, $u\in \C^n$, and assume that $Y$ has the
form \eqref{S12} then \eqref{S13} turns into the following  equations
for $A$ and $B$
\be
\frac{1}{2\ri}\big(e^{\ri q}Ae^{-\ri q}-e^{-\ri q}Ae^{\ri q}\big)
+\mu(uu^\dag-\1_n)=\0_n,
\label{S14}
\ee
and
\be
\frac{1}{2\ri}\big(e^{\ri q}Be^{\ri q}-e^{-\ri q}Be^{-\ri q}\big)
-\kappa e^{-2\ri q}-\mu uu^\dag+(\mu-\nu)\1_n=\0_n.
\label{S15}
\ee
Since $\mu \neq 0$, equation (\ref{S14}) implies that $\vert u_j \vert^2 =1$ for all $j=1,\ldots, n$.
Therefore we can apply a `residual' gauge transformation by an element
$(g_L, g_R) = (e^{\ri \xi(x)}, e^{\ri \xi(x)})$, with suitable $e^{\ri \xi(x)}\in Z$ (\ref{P28})
 to transform
$\upsilon_{\mu,\nu}^\ell(V)$ into $\upsilon_{\mu,\nu}^\ell(V_\R)$.
This amounts to setting $u_j=1$ for all $j=1,\ldots, n$.
After having done this,
we return to equations \eqref{S14} and \eqref{S15}. By writing
out the equations entry-wise, we obtain that the diagonal components of $A$ are
arbitrary imaginary numbers (which we denote by $\ri p_1,\ldots,\ri p_n$)
and we also obtain the following system of equations
\be
\begin{split}
A_{j,k}\sin(q_j-q_k)=-\mu=-B_{j,k}\sin(q_j+q_k),&\quad j\neq k,\\
B_{j,j}\sin(2q_j)=\nu+\kappa\cos(2q_j)-\ri\kappa\sin(2q_j),&\quad j,k=1,\ldots,n.
\end{split}
\label{S16}
\ee
So far we only knew that $q$ satisfies $\pi/2\geq q_1  \geq \ldots \geq q_n \geq 0$.
By virtue of  the conditions (\ref{S5}), the system  (\ref{S16}) can be solved if and only if
$\pi/2>q_1 >\cdots>q_n>0$. Substituting the unique solution for $A$ and $B$
back into (\ref{S12})
gives the formula $Y=Y(q,p)$ as displayed in (\ref{S2}).

The above arguments show that every gauge orbit in $P_0$ contains a point of $S$ (\ref{S6}),
 and  it is immediate by turning the equations backwards that every point of $S$ belongs to $P_0$.
By using that $q$ satisfies strict inequalities and that all components of $V_\R$
are non-zero, it is also readily seen  that no two different points of $S$ are gauge equivalent.
Moreover, the effectively acting symmetry group, which is given by
\be
(G_+ \times G_+)/\mathrm{U}(1)_{\diag}
\label{S17}
\ee
where $\mathrm{U}(1)$ contains the scalar unitary matrices,  acts \emph{freely} on $P_0$.

It follows from the above that $P_\red$ is a smooth manifold diffeomorphic to $M$.
Now the proof is finished by direct computation of the pull-back of the symplectic form
$\Omega$
of $P$ (\ref{P17}) onto the global cross-section $S$.
\end{proof}

Let us recall that the Abelian Poisson algebras $\fQ^1$ and $\fQ^2$ (\ref{P23})
consist of $(G_+ \times G_+)$-invariant functions  on $P$, and thus descend to
Abelian Poisson algebras on the reduced phase space $P_\red$.
In terms of the model $M\simeq S\simeq P_\red$, the Poisson algebra $\fQ^2_\red$ is
obviously generated by the functions
$(q,p)\mapsto \tr ((-\ri Y(q,p)))^m$ for $m=1,\ldots, N$.
It will be shown in the following section\footnote{In fact, we shall see
that $Y(q,p)$ is conjugate to
a diagonal matrix $\ri \Lambda$ of the form in equation (\ref{d7}).}
that these functions vanish identically for the odd integers,
and functionally independent generators of $\fQ^2_\red$ are provided by the functions
\be
H_k(q,p):= \frac{1}{4k}\tr(-\ri Y(q,p))^{2k},
\qquad\quad k=1,\ldots, n.
\label{S18}
\ee
The first of these functions reads
\be\begin{split}
H_1(q,p)=\frac{1}{4}\tr(-\ri Y(q,p))^2=&
\frac{1}{2}\sum_{j=1}^np_j^2
+\sum_{1\leq j<k\leq n}\bigg(\frac{\mu^2}{\sin^2(q_j-q_k)}
+\frac{\mu^2}{\sin^2(q_j+q_k)}\bigg)\\
&+\frac{1}{2}\sum_{j=1}^n\frac{\nu\kappa}{\sin^2(q_j)}
+\frac{1}{2}\sum_{j=1}^n\frac{(\nu-\kappa)^2}{\sin^2(2q_j)}.
\end{split}
\label{S19}
\ee
That is, upon the identification (\ref{I6}) it coincides with the Sutherland Hamiltonian (\ref{I1}).
This implies the Liouville integrability of the Hamiltonian (\ref{I1}).
Since its spectral invariants yield a commuting family of $n$ independent
functions in involution that include the Sutherland Hamiltonian,
the Hermitian matrix function $-\ri Y(q,p)$ (\ref{S2}) serves as a Lax matrix for
the Sutherland system $(M,\omega,H)$.

As for the reduced Abelian Poisson algebra $\fQ^1_\red$, we notice that
the cross-section $S$ permits to identify it with the Abelian Poisson algebra
of the smooth functions of the variables $q_1,\ldots, q_n$.
This is so since the level set $P_0$ lies completely in the `regular
part' of the phase space $P$, where the $G$-component $y$ of $(y,Y, \upsilon^\ell, \upsilon^r)$
is such that $Q(q)$ in its decomposition (\ref{P31}) satisfies strict inequalities
$\pi/2>q_1>\cdots>q_n>0$.
It is a  well-known fact
that in the regular part the components of $q$ are smooth (actually real-analytic)
functions
of $y$ (while globally they are only continuous functions).
To see that every smooth function depending on $q\in C_1$ is contained in $\fQ^1_\red$,
one may further use that  every  $(G_+\times G_+)$-invariant smooth
function on $P_0$ can be extended to an
invariant smooth function on $P$. Indeed, this holds since $G_+ \times G_+$ is compact and
$P_0 \subset P$ is a regular submanifold,
which itself follows from the free action property established in the course of the proof
of Theorem 3.1.

We can summarize the outcome of the foregoing discussion as follows.
Below, the generators of Poisson algebras are understood in the functional sense, i.e.,
if some $f_1,\ldots, f_n$ are generators then all smooth functions of them belong
to the Poisson algebra.

\medskip
\noindent \bf\sffamily Corollary 3.2. \it
By using the model $(M,\omega)$ of the reduced phase space $(P_\red, \Omega_\red)$
provided by Theorem 3.1,
the Abelian Poisson algebra $\fQ^2_\red$ (2.23) can be identified with the Poisson algebra
generated by the
spectral invariants (3.18) of the `Sutherland Lax matrix' $-\ri Y(q,p)$ (\ref{S2}),
which according to (\ref{S19}) include
the many-body Hamiltonian $H(q,p)$ (\ref{I1}), and $\fQ^1_\red$ can be identified
with the algebra generated by the corresponding position variables $q_i$ $(i=1,\ldots, n)$.
\rm

\section{The dual picture}
\setcounter{equation}{0}

It follows from the group-theoretic results quoted in Section 2.2 that the Abelian
Poisson algebra $\fQ^1$ is generated by
the functions
\be
\tilde\cH_k(y,Y,\upsilon^\ell,\upsilon^r):=\frac{(-1)^k}{2k}\tr\big(y^{-1}CyC\big)^k,
\quad k=1,\ldots, n,
\label{d1}
\ee
and thus the unitary and Hermitian matrix
\be
L:=-y^{-1}CyC
\label{d2}
\ee
serves as an `unreduced Lax matrix'.  It is readily seen in the Sutherland gauge
 (\ref{S6})
that these $n$ functions remain functionally independent after reduction.
Here, we shall prove that the evaluation of the invariant function $\tilde\cH_1$ in
another gauge reproduces the dual Hamiltonian (\ref{I4}). The reduction
of the matrix function $L$ will provide a Lax matrix for the corresponding integrable system.
Before turning to details, we  advance the group-theoretic interpretation of the dual position
variable $\lambda$
that features in the Hamiltonian (\ref{I4}), and sketch the plan of this section.

To begin, recall that on the constraint surface $Y= Y_- - \ri \kappa C$, and
for any $Y_-\in\cG_-$ there is an element $g_R\in G_+$ such that
\be
g_R^{-1}Y_- g_R^{\phantom{1}}
=\diag(\ri d_1,\ldots,\ri d_n,-\ri d_1,\ldots,-\ri d_n)=\ri D\in\cA
\quad\text{with}\quad d_1\geq\cdots\geq d_n\geq 0.
\label{d3}
\ee
Then introduce the real matrix
$\blambda=\diag(\lambda_1,\ldots,\lambda_n)$ whose diagonal
components are\footnote{From now on we frequently use the notations
$\N_n:=\{1,\ldots,n\}$ and $\N_N:=\{1,\ldots,N\}$.}
\be
\lambda_j:=\sqrt{d_j^2+\kappa^2},\quad j\in\N_n.
\label{d4}
\ee
One can diagonalize the matrix $D-\kappa C$ by conjugation with the unitary matrix
\be
h(\lambda)=\begin{bmatrix}
\alpha(\blambda)&\beta(\blambda)\\
-\beta(\blambda)&\alpha(\blambda)
\end{bmatrix},
\label{d5}
\ee
where the real functions $\alpha(x),\beta(x)$ are defined on the interval
$[|\kappa|,\infty)\subset\R$ by the formulae
\be
\alpha(x)=\frac{\sqrt{x+\sqrt{x^2-\kappa^2}}}{\sqrt{2x}},\quad
\beta(x)=\kappa\frac{1}{\sqrt{2x}}\frac{1}{\sqrt{x+\sqrt{x^2-\kappa^2}}},
\label{d6}
\ee
at least if $\kappa \neq 0$. If $\kappa=0$, then we set $\alpha(x)=1$ and $\beta(x)=0$.
Indeed, it is easy to check that
\be
h(\lambda)\Lambda h(\lambda)^{-1}=D-\kappa C
\quad\text{with}\quad
\Lambda=\diag(\lambda_1,\ldots,\lambda_n,-\lambda_1,\ldots,-\lambda_n).
\label{d7}
\ee
Note that $h(\lambda)$ belongs to the subset $G_-$  of $G$ (\ref{P10}).

The above diagonalization procedure can be used to define the map
\be
\fL\colon P_0\to \R^n,\quad
(y,Y,\upsilon^\ell,\upsilon^r)\mapsto\lambda.
\label{d8}
\ee
This is clearly a continuous map, which descends to a continuous map
$\fL_\red: P_\red \to \R^n$.  One readily sees also that
these maps are smooth (even real-analytic)
on  the open submanifolds
$P_0^\reg\subset P_0$ and $P_\red^\reg\subset P_\red$, where the
$N$ eigenvalues of $Y_-$ are pairwise different.

The image of the
constraint surface $P_0$ under the map $\fL$ will turn out to be  the closure of the domain
\be
C_2=\bigg\{\lambda\in\R^n\bigg|
\begin{matrix}\lambda_a-\lambda_{a+1}>2\mu,\\
(a=1,\ldots,n-1)\end{matrix}
\quad\text{and}\quad
\lambda_n>\nu \bigg\}.
\label{d9}
\ee
By solving the constrains through the diagonalization of $Y$,
we shall construct a model of the open
submanifold of $P_\red$ corresponding to the open submanifold
$\fL^{-1}(C_2) \subset P_0$.
This model will be symplectomorphic to the semi-global phase-space
$C_2 \times \T^n$ of the  dual Hamiltonian (\ref{I4}).

In Subsection 4.1, we present the construction of the aforementioned model
of $\fL_\red^{-1}(C_2)\subset P_\red$. The proof that also enlightens the
origin of the construction given in Subsection 4.2. In Subsection 4.3 we
demonstrate that $\fL_\red^{-1}(C_2)$ is a dense subset of $P_\red$ and
finally, in Subsection 4.4 we present the global characterization of the
dual model of $P_\red$.

Many of the local formulae that appear in this section have analogues
in \cite{P1,P2,P3}, which inspired our considerations. However, the
global structure is different.

\subsection{The dual model of the open subset $\fL^{-1}_\red(C_2) \subset P_\red$}

We first prepare some functions on $C_2 \times \T^n$. Denoting
the elements of this domain as pairs
\be
(\lambda,e^{\ri \vartheta})
\quad\text{with}\quad
\lambda=(\lambda_1,\ldots,\lambda_n)\in C_2,\quad
e^{\ri\vartheta}=(e^{\ri\vartheta_1},\ldots,e^{\ri\vartheta_n})\in\T^n,
\label{d10}
\ee
we let
\bea
&&\phantom{+c} f_{c} :=\bigg[1-\frac{\nu}{\lambda_c}\bigg]^{\frac{1}{2}}
\prod_{\substack{a=1\\(a\neq c)}}^n
\bigg[1-\frac{2\mu}{\lambda_c-\lambda_a}\bigg]^{\frac{1}{2}}
\bigg[1-\frac{2\mu}{\lambda_c+\lambda_a}\bigg]^{\frac{1}{2}}, \quad \forall c\in \N_n,
\nonumber\\
&&f_{n+c}:=e^{\ri \vartheta_c} \bigg[1+\frac{\nu}{\lambda_c}\bigg]^{\frac{1}{2}}
\prod_{\substack{a=1\\(a\neq c)}}^n
\bigg[1+\frac{2\mu}{\lambda_c-\lambda_a}\bigg]^{\frac{1}{2}}
\bigg[1+\frac{2\mu}{\lambda_c+\lambda_a}\bigg]^{\frac{1}{2}}.
\label{d11}
\eea
For $\lambda \in C_2$ (\ref{d9}), all factors under the square roots are positive.
Using the column vector $f:= (f_1,\ldots, f_{2n})^\top$ together with
 $\Lambda_c =\lambda_c$ and $\Lambda_{c+n} = -\lambda_c$  for $c\in \N_n$,
we define the $N\times N$ matrices $\check A(\lambda, \vartheta)$ and
$B(\lambda, \vartheta)$ by
\be
\check A_{j,k}=\frac{2\mu f_j\overline{(Cf)}_k-
2(\mu-\nu)C_{j,k}}{2\mu+\Lambda_k-\Lambda_j},\quad
j,k\in\N_N,
\label{d12}
\ee
and
\be
B(\lambda,\vartheta):=-\big(h(\lambda)
\check A(\lambda,\vartheta)h(\lambda)\big)^\dagger.
\label{d13}
\ee
We shall see that these are unitary matrices from $G_-\subset G$ (\ref{P10}).
Then we write $B$ in the form
\be
B=\eta e^{2\ri Q(q)}\eta^{-1}
\label{d14}
\ee
with some $\eta\in G_+$ and unique $q= q(\lambda,\vartheta)$ subject to
(\ref{P30}). (It turns out that $q(\lambda,\vartheta)\in C_1$ (\ref{I2})
and thus $\eta$ is unique up to replacements $\eta\to\eta\zeta$ with
arbitrary $\zeta \in Z$ (\ref{P28}).) Relying on (\ref{d14}), we set
\be
y(\lambda,\vartheta):=\eta e^{\ri Q(q(\lambda,\vartheta))}\eta^{-1}
\label{d15}
\ee
and introduce the vector $V(\lambda,\vartheta)\in\C^N$ by
\be
V(\lambda,\vartheta):=y(\lambda,\vartheta)h(\lambda)f(\lambda,\vartheta).
\label{d16}
\ee
It will be shown that $V+CV=0$ and $|V|^2=N$, which ensures that
$\upsilon^\ell_{\mu,\nu}(V)\in\cO^\ell$ (\ref{P14}).

Note that  $\check A$, $y$ and $V$  given above
depend on $\vartheta$ only through
$e^{\ri \vartheta}$ and
are $C^\infty$ functions on $C_2\times\T^n$.
It should be remarked that
although the matrix element $\check A_{n,2n}$
(\ref{d12}) has an apparent
singularity at $\lambda_n=\mu$, the zero of
the denominator cancels. Thus $\check A$ extends by continuity
to $\lambda_n=\mu$ and remains smooth there, which then also implies the
smoothness of $y$ and $V$.

\medskip
\noindent \bf\sffamily Theorem 4.1. \it
By using the above notations, consider the set
\be
\tilde S^0:=\{(y(\lambda,\vartheta),\ri h(\lambda)\Lambda(\lambda)h(\lambda)^{-1},
\upsilon^\ell_{\mu,\nu}(V(\lambda,\vartheta)),
\upsilon^r)\mid(\lambda,e^{\ri \vartheta})\in
C_2\times\T^n\}.
\label{d17}
\ee
This set is contained in the constraint surface $P_0=J^{-1}(0)$ and it
provides a cross-section
for the $G_+\times G_+$-action restricted to $\fL^{-1}(C_2)\subset P_0$.
In particular, $C_2\subset\fL(P_0)$ and
$\tilde S^0$ intersects every gauge orbit in $\fL^{-1}(C_2)$ precisely in one point.
Since the elements of $\tilde S^0$ are parametrized by $C_2\times\T^n$ in
a smooth and bijective
manner, we obtain the identifications
\be
\fL^{-1}_\red(C_2)\simeq\tilde S^0\simeq C_2\times\T^n.
\label{d18}
\ee
Letting $\tilde \sigma_0: \tilde S^0 \to P$ denote the tautological injection,
the pull-backs of the symplectic form $\Omega$ (\ref{P17}) and the function
$\tilde\cH_1$ (\ref{d1}) obey
\be
\tilde \sigma_0^*(\Omega)=\sum_{c=1}^n\rd\lambda_c\wedge\rd\vartheta_c,
\qquad
(\tilde \cH_1 \circ \tilde \sigma_0)(\lambda,\vartheta)
=\frac{1}{2}\tr\big(h(\lambda)\check A(\lambda,\vartheta)h(\lambda)\big)
=\tilde H^0(\lambda,\vartheta)
\label{d19}
\ee
with the RSvD  type Hamiltonian $\tilde H^0$ in (\ref{I4}).
Consequently, the Hamiltonian reduction of the system $(P,\Omega,\tilde\cH_1)$
followed by restriction to
the open submanifold $\fL_\red^{-1}(C_2)\subset P_\red$
reproduces the system $(\tilde M^0,\tilde\omega^0,\tilde H^0)$ defined in the Introduction.

\rm

\medskip
\noindent \bf\sffamily  Remark 4.2. \rm
Referring to (\ref{d2}), we have the Lax matrix
\be
L(y(\lambda, \vartheta))=h(\lambda)\check A(\lambda,\vartheta)h(\lambda).
\label{d20}
\ee
Later we shall also prove that $\fL_{\red}^{-1}(C_2)$ is a dense subset of $P_\red$,
whereby the reduction of
$(P,\Omega,\tilde\cH_1)$ may be viewed as a completion of
$(\tilde M^0,\tilde\omega^0,\tilde H^0)$.

\subsection{Proof of Theorem 4.1}

The proof will emerge from a series of lemmas.
 Our immediate aim is to
construct gauge invariant functions that will be used for
parametrizing the orbits of $G_+\times G_+$ in (an open submanifold of) $P_0$.
For introducing gauge invariants we can restrict ourselves to the submanifold
$P_1\subset P_0$ where $Y$ in $(y,Y,\upsilon^\ell,\upsilon^r)$ has the form
\be
Y=h(\lambda)\ri\Lambda(\lambda)h(\lambda)^{-1}
\label{d21}
\ee
with some $\lambda\in\R^n$ for which
\be
\lambda_1 \geq\cdots\geq\lambda_n\geq|\kappa|.
\label{d22}
\ee
Indeed, every element of $P_0$ can be gauge transformed into $P_1$.
It will be advantageous to further restrict attention to
$P_1^\reg\subset P_1$ where we have
\be
\lambda_1 >\cdots>\lambda_n>|\kappa|.
\label{d23}
\ee
The residual gauge transformations that map $P_1^\reg$ to
itself belong  to the group $G_+\times Z < G_+\times G_+$ with $Z$
defined in \eqref{P28}.
Since $\upsilon^r$ is constant and $\upsilon^\ell=\upsilon^\ell_{\mu,\nu}(V)$, we may
label the elements of $P_1$ by triples $(y,Y,V)$, with the understanding that $V$
matters up to phase.
Then the gauge action of $(g_L,\zeta)\in G_+ \times Z$ operates by
\be
(y,V)\mapsto(g_Ly\zeta^{-1},g_LV),
\label{d24}
\ee
while $Y$ is already invariant. Now we can factor out the residual $G_+$-action by
introducing the $G_-$-valued function
\be
\check A(y,Y,V):=h(\lambda)^{-1}L(y)h(\lambda)^{-1}
\label{d25}
\ee
and the $\C^N$-valued function
\be
F(y,Y,V):=h(\lambda)^{-1}y^{-1}V.
\label{d26}
\ee
Here $\lambda=\fL(y,Y,V)$, which means that (\ref{d21}) holds,
and we used $L(y)$ in (\ref{d2}).
Like $V$, $F$ is defined only up to a $\UN(1)$ phase.
We obtain the transformation rules
\be
\check A(g_Ly\zeta^{-1},Y,g_L V)=\zeta\check A(y,Y,V)\zeta^{-1},
\label{d27}
\ee
\be
F(g_Ly\zeta^{-1},Y,g_L V)=\zeta F(y,Y,V),
\label{d28}
\ee
and therefore the functions
\be
\cF_k(y,Y,V):=|F_k(y,Y,V)|^2,\quad
k=1,\ldots,N
\label{d29}
\ee
are well-defined, gauge invariant, smooth functions on $P_1^\reg$.
They represent $(G_+\times G_+)$-invariant smooth functions on $P_0^\reg$.
We shall see shortly that the functions $\cF_k$ depend only on
$\lambda=\fL(y,Y,V)$ and shall derive explicit formulae for this dependence.
Then the non-negativity of $\cF_k$ will be used to gain information about
the set $\fL(P_0)$ of $\lambda$ values that actually occurs.

Before turning to the inspection of the functions $\cF_k$,
we present a crucial lemma.

\medskip
\noindent \bf\sffamily Lemma 4.3. \it
Fix $\lambda\in\R^n$ subject to (\ref{d22}) and set
$\Lambda:=\diag(\lambda,-\lambda)$ and $Y:=h(\lambda)\ri\Lambda h(\lambda)^{-1}$.
If $y\in G$ and $\upsilon^\ell_{\mu,\nu}(V)\in\cO^\ell$ solve the momentum map
constraint given according to the first equation in (\ref{S8}) by
\be
yYy^{-1}+CyYy^{-1}C+2\upsilon^\ell_{\mu,\nu}(V)=0,
\label{d30}
\ee
then $\check A\in G_-$ and $F\in\C^N$ defined by (\ref{d25}) and (\ref{d26})
solve the following equation:
\be
2\mu\check{A}+\check{A}\Lambda-\Lambda\check{A}
=2\mu F(CF)^\dag-2(\mu-\nu)C.
\label{d31}
\ee
Conversely, for any $\check A\in G_-$, $F\in\C^N$
that satisfy $|F|^2 =N$ and equation (\ref{d31}),
pick $y\in G$ such that $L(y)=h(\lambda)\check Ah(\lambda)$
and define $V:=yh(\lambda)F$. Then $CV+V=0$ and $(y,Y,\upsilon^\ell_{\mu,\nu}(V))$
solve the momentum map constraint (\ref{d30}).
\rm

\begin{proof}
If eq.~(\ref{d30}) holds, then we multiply it by $h(\lambda)^{-1}y^{-1}$ on
the left and by $CyCh(\lambda)^{-1}$ on the right.
Using (\ref{S13}), with $CV+V=0$ and $|V|^2=N$, and the notations (\ref{d25}) and (\ref{d26}),
this immediately gives (\ref{d31}).
Conversely, suppose that (\ref{d31}) holds for some $\check A\in G_-$ and $F\in\C^N$ with
$|F|^2=N$.  Since $h(\lambda)\check Ah(\lambda)$ belongs to $G_-$, there exists $y\in G$
such that
\be
h(\lambda)\check Ah(\lambda)=L(y).
\label{d32}
\ee
Such $y$ is unique up to left-multiplication by an arbitrary element of $G_+$
(whereby one may bring $y$ into $G_-$ if one wishes to do so).
Picking $y$ according to (\ref{d32}), and then setting
\be
V:=yh(\lambda)F,
\label{d33}
\ee
it is an elementary matter to show that (\ref{d31}) implies the following equation:
\be
yYy^{-1}+CyYy^{-1}C+2\ri\mu(-V(CV)^\dagger-\1_N)+2\ri(\mu-\nu)C=0.
\label{d34}
\ee
It is a consequence of this equation that
\be
(V(CV)^\dagger)^\dagger=(CV)V^\dagger=V(CV)^\dagger.
\label{d35}
\ee
This entails that $CV=\alpha V$ for some $\alpha\in\UN(1)$.
Then $V^\dagger=\alpha(CV)^\dagger$ also holds, and thus we must have $\alpha^2=1$.
Hence $\alpha$ is either $+1$ or $-1$. Taking the trace of the equality (\ref{d34}), and using that
$|V|^2=N$ on account of $|F|^2=N$, we obtain that
$\alpha =-1$, i.e., $CV+V=0$. This means that equation (\ref{d34}) reproduces (\ref{d30}).
\end{proof}

To make progress, now we restrict our
attention to the subset of $P_1^\reg$ where the eigenvalue-parameter $\lambda$
of $Y$ verifies in addition to (\ref{d23}) also the conditions
\be
|\lambda_a\pm\lambda_b|\neq 2\mu
\quad\text{and}\quad
(\lambda_a-\nu)(\lambda_a-\vert 2\mu-\nu\vert)\neq 0,
\quad
\forall a,b\in\N_n.
\label{d36}
\ee
We call such $\lambda$ values `strongly regular', and  let
$P_1^{\vreg}\subset P_1$ and $P_0^{\vreg}\subset P_0$ denote the corresponding
open subsets.
Later we shall prove that $P_0^\vreg$ is \emph{dense} in $P_0$.
The above conditions will enable us to perform calculations
that will lead to a description of a dense subset of the reduced phase space.
They ensure that we never divide by zero in relevant steps of our arguments.
The first such step is the derivation of the  following consequence
of equation (\ref{d31}).

\medskip
\noindent \bf\sffamily Lemma 4.4. \it
The restriction of the matrix function $\check A$ (\ref{d25}) to $P_1^{\vreg}$
has the form
\be
\check A_{j,k}=\frac{2\mu F_j\overline{(C F)}_k-
2(\mu-\nu)C_{j,k}}{2\mu+\Lambda_k-\Lambda_j},\quad
j,k\in\N_N,
\label{d37}
\ee
where $F\in\C^N$ satisfies
$|F|^2=N$  and $\Lambda=\diag(\lambda,-\lambda)$
varies on $P_1^\vreg$ according to (\ref{d21}).
\rm

\medskip
\noindent \bf\sffamily Lemma 4.5. \it
For any strongly regular $\lambda$ and $a\in\N_n$ define
\be
w_a:=\prod_{\substack{b=1\\(b\neq a)}}^n
\frac{(\lambda_a-\lambda_b)(\lambda_a+\lambda_b)}
{(2\mu-(\lambda_a-\lambda_b))(2\mu-(\lambda_a+\lambda_b))},\quad
w_{a+n}:=\prod_{\substack{b=1\\(b\neq a)}}^n
\frac{(\lambda_a-\lambda_b)(\lambda_a+\lambda_b)}
{(2\mu+\lambda_a-\lambda_b)(2\mu+\lambda_a+\lambda_b)},
\label{d38}
\ee
and set $W_k:=w_k\cF_k$ with $\cF_k=|F_k|^2$. Then the unitarity of the matrix
$\check A$ as given by (\ref{d37}) implies the following system of equations for
the pairs of functions $W_c$ and $W_{c+n}$ for any $c\in\N_n$:
\be
(\mu+\lambda_c)W_c+(\mu-\lambda_c)W_{n+c}-2(\mu-\nu)=0,
\label{d39}
\ee
\be
\lambda_c^2W_cW_{n+c}
-\mu(\mu-\nu)(W_c+W_{n+c})
+(\mu-\nu)^2+\mu^2-\lambda_c^2=0.
\label{d40}
\ee
For fixed $c\in\N_n$ and strongly regular $\lambda$, this system of equations
admits two solutions, which are given by
\be
(W_c,W_{n+c})=(W_c^+,W_{n+c}^+)=(w_c\cF_c^+,w_{c+n}\cF_{c+n}^+)
=(1+\frac{\nu}{\lambda_c},1-\frac{\nu}{\lambda_c}),
\label{d41}
\ee
and by
\be
(W_c,W_{n+c})=(W_c^-,W_{n+c}^-)=(w_c\cF_c^-,w_{c+n}\cF_{c+n}^-)
=(-1+\frac{2\mu-\nu}{\lambda_c},-1-\frac{2\mu-\nu}{\lambda_c}).
\label{d42}
\ee
The functions $\cF_k^\pm$ satisfy the identities
\be
\sum_{k=1}^N\cF_k^+(\lambda)=N
\quad\text{and}\quad
\sum_{k=1}^N\cF_k^-(\lambda)=-N.
\label{d43}
\ee
\rm
\begin{proof}
The derivation of equations (\ref{d39}), (\ref{d40})
follows a similar derivation due to Pusztai \cite{P1},
and is summarized in the appendix. We then solve the linear equation (\ref{d39}) say for $W_{c+n}$ and substitute
it into (\ref{d40}). This gives a quadratic equation for $W_c$ whose two solutions we can write down.
We note that the derivation of the equations (\ref{d39}) and (\ref{d40})
presented in the appendix utilizes the full set of the conditions (\ref{d36}).

To verify the identities (\ref{d43}), we first extend $\lambda$ to
vary in the open subset of $\C^n$ subject to the conditions
$\lambda_a^2\neq\lambda_b^2$ and
$\lambda_c \neq 0$, and then consider the sums
that appear in (\ref{d43}) as
functions of a chosen
component of $\lambda$ with the other components fixed.
These explicitly given sums are meromorphic functions having only first order poles,
and one may check that all residues at the apparent poles vanish.
Hence the sums are constant over $\C^n$, and
the values of the constants can be established by looking at
a suitable asymptotic limit in the domain $C_2$ (\ref{d9}), whereby all $w_k$ tend to 1
and the pre-factors in (\ref{d41}) and (\ref{d42}) tend to $1$ and $-1$, respectively.
\end{proof}

Observe that neither any $w_k$ nor any $\cF_k^\pm$ ($k\in\N_N$) can vanish if
$\lambda$ is strongly regular.
We know that the value of $\cF_k$ (\ref{d29}) is uniquely defined at every point of $P_1^\reg$.
Therefore only one of the solutions  $(\cF_c^\pm,\cF_{c+n}^\pm)$ can be acceptable at any
$\lambda\in\fL(P_1^\vreg)$.
The identities in (\ref{d43}) and analyticity arguments strongly suggest that the acceptable solutions
are provided by $\cF_k^+$.
The first statement of the following lemma confirms that this is the case
for  $\lambda\in C_2$ (\ref{d9}).

\medskip
\noindent \bf\sffamily Lemma 4.6. \it
The formulae (\ref{d41}) and (\ref{d42}) can be used to define $\cF_k^\pm$ as smooth real functions
on the domain $C_2$, and none of these functions vanishes at any $\lambda\in C_2$.
Then for any $\lambda\in C_2$ and $c\in\N_n$ at least one out of $\cF_c^-$ and $\cF^-_{c+n}$ is negative,
while $\cF_{k}^+>0$ for all $k\in\N_N$.
Hence for  $\lambda\in C_2\cap\fL(P_0)$
only $\cF_k^+(\lambda)$ can give the value of the function $\cF_k$ as defined in
(\ref{d29}).
Taking any
$\lambda\in C_2$ and any $F\in\C^N$ satisfying $|F_k|^2=\cF_k^+(\lambda)$,
the formula (\ref{d37})
yields a unitary matrix that belongs to $G_-$ (\ref{P10}).
This matrix $\check A$ and vector $F\in\C^N$ solve equation (\ref{d31}).
\rm

\begin{proof}
It is easily seen that $w_k(\lambda)>0$ for all $\lambda\in C_2$ and $k\in\N_N$.
The statement about the negativity of either $\cF_c^-$ or $\cF^-_{c+n}$ thus follows from
the identity $W_c^-+W_{n+c}^-=-2$. The positivity of
$\cF_k^+$ is easily checked.
It is also readily verified that $\check A^\dagger=C\check AC$,
which entails that $\check A \in G_-$ once we know that $\check A$ is unitary.
For  $\lambda \in C_2$ and $|F_k|^2=\cF_k^+(\lambda)$, the unitarity of $\check A$ (\ref{d37})
can be shown by almost verbatim adaptation of the arguments proving Proposition 6 in \cite{P2}.

If $\lambda\in C_2$ is such that the denominators in (\ref{d37}) do not vanish,
then the formula (\ref{d37}) is plainly
equivalent to (\ref{d31}).
Observe that only those elements $\lambda\in C_2$ for which $\lambda_n=\mu$ fail to satisfy
this condition.
At such $\lambda$ the matrix element
$\check A_{n,2n}$ has an apparent `first order pole', but one can check
by inspection of the formula (\ref{d12}) that  $\check A_{n,2n}$ actually
remains finite and smooth even at such exceptional points,
and thus solves also (\ref{d31}) because of continuity.
\end{proof}

\medskip

Before presenting the proof of Theorem 4.1,
note that  at the point of $\tilde S^0$ labeled by
$(\lambda,e^{\ri\vartheta})$
the value of the function $F$ (\ref{d26}) is equal to
$f(\lambda,e^{\ri\vartheta})$ given in (\ref{d11}).
\medskip

\begin{proof}[Proof of Theorem 4.1]
It follows from Lemma 4.3 and Lemma 4.6 that $\tilde S^0$ is a subset of
$P_1^\reg$ and
$\fL(\tilde S^0)= C_2$.  Taking into account Theorem 3.1, this implies that
$y(\lambda,\vartheta)$ (\ref{d15}) and $V(\lambda, \vartheta)$  (\ref{d16})
are well-defined smooth functions on $C_2 \times \T^n$.
We next show that $\tilde S^0$ is a cross-section for the
residual gauge action on $\fL^{-1}(C_2) \cap P_1$.
To do this, pick an arbitrary element
\be
(\tilde y, h(\lambda) \ri  \Lambda h(\lambda)^{-1},
\upsilon_{\mu,\nu}^\ell(\tilde V), \upsilon^r)
\in \fL^{-1}(C_2) \cap P_1.
\label{d44}
\ee
Because $\cF_k(\lambda)\neq 0$, we can find a unique element
$e^{\ri \vartheta}\in \T^n$ and
an element $\zeta \in Z$ (\ref{P28}) (which is unique up to scalar multiple)
such that
\be
F_k(\tilde y \zeta^{-1}, h(\lambda) \ri  \Lambda h(\lambda)^{-1},
\tilde V) = f_k(\lambda, e^{\ri \vartheta}), \qquad \forall k\in \N_N,
\label{d45}
\ee
up to a $k$-independent phase.
We then see from (\ref{d31}) that
$L(\tilde y \zeta^{-1})=L(y(\lambda,\vartheta))$, which in turn
implies the existence of some (unique after $\zeta$ was chosen)
$\eta_+ \in G_+$ for which
\be
\eta_+ \tilde y \zeta^{-1} = y(\lambda, \vartheta).
\label{d46}
\ee
Using also that $\zeta^{-1} h(\lambda) \zeta = h(\lambda)$,
we conclude from the last two equations that
\be
\eta_+\tilde V
=\eta_+\tilde y h(\lambda)F(\tilde y,h(\lambda)\ri\Lambda h(\lambda)^{-1},
\tilde V)
=y(\lambda,\vartheta)h(\lambda)f(\lambda,\vartheta)
=V(\lambda,e^{\ri\vartheta}).
\label{d47}
\ee
Thus we have shown that the element (\ref{d44}) can be gauge transformed
into a point of $\tilde S^0$, and
this point is uniquely determined since (\ref{d45}) fixes $e^{\ri\vartheta}$
uniquely.
In other words, $\tilde S^0$ intersects every  orbit of the residual
gauge action on $\fL^{-1}(C_2) \cap P_1$ in precisely one point.

The map from $C_2$ into $P$, given by the parametrization of $\tilde S^0$,
 is obviously smooth, and hence we obtain the
identifications
\be
C_2 \simeq\tilde S^0\simeq(\fL^{-1}(C_2)\cap P_1)/(G_+\times Z)
\simeq\fL^{-1}(C_2)/(G_+\times G_+)=\fL_\red^{-1}(C_2).
\label{d48}
\ee
To establish the formula (\ref{d19}) of the reduced symplectic structure,
we proceed as follows.
We define $G_+ \times G_+$ invariant real functions on $P$ by
\be
\varphi_m(y,Y,V):=\dfrac{1}{m}\Re\big(\tr(Y^m)\big),
\quad m\in\N,
\label{d49}
\ee
and
\be
\chi_k(y,Y,\upsilon):=\Re\big(\tr(Y^ky^{-1} V V^\dagger yC)\big),
\quad k\in\N \cup\{0\}.
\label{d50}
\ee
The restrictions of these functions to $\tilde S^0$ are the respective
functions $\varphi_m^\red$ and $\chi_k^\red$:
\be
\varphi_m^\red(\lambda,\vartheta)=
\begin{cases}
0,&\text{if }m\text{ is odd},\\
\displaystyle(-1)^{\tfrac{m}{2}}\frac{2}{m}\sum_{j=1}^n\lambda_j^m,
&\text{if }m\text{ is even},
\end{cases}
\label{d51}
\ee
and
\be
\chi_k^\red(\lambda,\vartheta)=
\begin{cases}
\displaystyle - 2 (-1)^{\tfrac{k-1}{2}} \sum_{j=1}^n\lambda_j^k
\bigg[1-\frac{\kappa^2}{\lambda_j^2}\bigg]^{\frac{1}{2}}
X_j\sin(\vartheta_j),&\text{if }k\text{ is odd},\\
\displaystyle 2(-1)^{\tfrac{k}{2}} \sum_{j=1}^n\lambda_j^k
\bigg[1-\frac{\kappa^2}{\lambda_j^2}\bigg]^{\frac{1}{2}}
X_j\cos(\vartheta_j)
-\kappa\lambda_j^{k-1}\big(\cF_j-\cF_{n+j}\big),&\text{if }k\text{ is even},
\end{cases}
\label{d52}
\ee
where $X_j=\sqrt{\cF_j \cF_{n+j}}$.
Then we calculate the pairwise Poisson brackets of the set of functions
$\varphi_m$, $\chi_k$
on $P$ and restrict the results to $\tilde S^0$. This must coincide with
the results
of the direct calculation of the Poisson brackets of the reduced functions
$\varphi_m^\red$, $\chi_k^\red$ based on the pull-back of the symplectic
form $\Omega$ onto $\tilde S^0 \subset  P$. Inspection shows that the
required equalities
hold if and only if we have the formula in (\ref{d19}) for the pull-back
in question. This reasoning is very similar to that used in \cite{P2} to find
the corresponding reduced symplectic form.
Since the underlying calculations are straightforward,
although rather laborious,
we here omit the details.
As for the formula for the restriction of $\tilde \cH_1$ to $\tilde S^0$
displayed in (\ref{d19}),
this is a matter of direct verification.
\end{proof}

\subsection{Density properties}

So far we dealt with the open subset $\fL_\red^{-1}(C_2)$ of the reduced
phase space. Here we show that Theorem 4.1 contains `almost all'
information about the dual system
since $\fL_\red^{-1}(C_2)\subset P_\red$ is a \emph{dense} subset.
This key result will  be proved by combining two lemmas.

\medskip
\noindent \bf\sffamily Lemma 4.7. \it
The subset $P_0^\vreg \subset P_0$ of the constraint surface where the
range of the eigenvalue map $\fL$ (\ref{d8}) satisfies the conditions
\eqref{d23} and \eqref{d36} is dense.
\rm

\begin{proof}
Let us first of all note that $P_0$ is a connected regular analytic
submanifold of $P$.
In fact, it is a regular (embedded) analytic submanifold of the analytic
manifold $P$ since the momentum map is analytic and zero is its regular
value (because the effectively acting gauge group (3.17) acts freely on
$P_0$). The connectedness follows from Theorem 3.1, which implies that
$P_0$ is diffeomorphic to the product of $S$ (3.6) and the group (3.17),
and both are connected.

For any $Y\in \cG$ denote by $\{\ri \Lambda_a\}_{a=1}^N$ the set of its
eigenvalues counted with multiplicities.
Then the following formulae
\be
\cR(y,Y,V):=\prod_{\substack{a,b=1\\(a\neq b)}}^N(\Lambda_a-\Lambda_b)
\prod_{a=1}^N(\Lambda_a^2-\kappa^2),
\label{G1}
\ee
\be
\cS(y,Y,V):=\prod_{\substack{a,b=1\\(a\neq b)}}^N[(\Lambda_a-\Lambda_b)^2-4\mu^2]
\prod_{a=1}^N\left[(\Lambda_a^2-\mu^2)(\Lambda_a^2-\nu^2)(\Lambda_a^2-(2\mu-\nu)^2)
\right].
\label{G2}
\ee
define analytic functions on $P_0$.
Indeed, $\cR$ and $\cS$  are symmetric polynomials in the eigenvalues
of $Y$, and hence can be expressed as polynomials in the coefficients
of the characteristic
polynomial of $Y$, which are polynomials in the matrix elements of $Y$.
The product $\cR \cS$ is also an analytic function on $P_0$, and
the subset $P_0^\vreg$, which we considered in Subsection 4.2, can be
characterized as
\be
P_0^\vreg = \{ x\in P_0 \mid \cR(x) \cS(x) \neq 0\}.
\label{G3}
\ee
It is clear from Theorem 4.1 that $\cR \cS$ does not vanish identically
on $P_0$.
Since the zero set of a non-zero analytic function on a connected analytic
manifold cannot
contain any open set, equation (\ref{G3}) implies that
$P_0^\vreg$  is a dense subset of $P_0$.
\end{proof}
\medskip

Let $\overline{C}_2$ be the closure of the domain $C_2 \subset \R^n$.
Eventually, it will turn out that $\fL(P_0) = \overline{C}_2$.
For now, we wish to prove the following.

\medskip
\noindent\bf\sffamily Lemma 4.8. \it
For every boundary point $\lambda^0\in\partial\overline{C}_2 $ there
exist an open ball $B(\lambda^0)\subset\R^n$ around $\lambda^0$ that
does not contain any strongly regular $\lambda$ which lies outside
$\overline{C}_2$ and belongs to $\fL(P_0)$.
\rm

\begin{proof}
We start by noticing that for any boundary point
$\lambda^0\in\partial\overline{C}_2$
there is a ball $B(\lambda^0)$ centered at $\lambda^0$ such
that any strongly regular $\lambda\in B(\lambda^0)\setminus\overline{C}_2$
is subject to either of the following: (i) there is an index
$a\in\{1,\ldots,n-1\}$ such that
\be
\lambda_a-\lambda_{a+1}<2\mu
\quad\text{and}\quad
\lambda_b-\lambda_{b+1}>2\mu
\quad\forall\;b<a,
\label{D.6}
\ee
or (ii) we have
\be
\lambda_a-\lambda_{a+1}>2\mu,
\quad
a=1,\ldots,n-1
\quad\text{and}\quad
 \lambda_n<\nu.
\label{D.7}
\ee

Let us consider a strongly regular $\lambda\in B(\lambda^0)$
that falls into case (i) \eqref{D.6} and is so close to $C_2$
that we still have
\be
\lambda_k-\lambda_{k+1}>\mu,\quad
\forall\;k\in\{1,\ldots,n-1\}.
\label{D.8}
\ee
It then follows that
\be
\lambda_a-\lambda_b>2\mu,\quad\forall\;b>a+1,
\label{D.9}
\ee
and
\be
\lambda_a+\lambda_b>2\mu,\quad\forall\;b\in\{1,\ldots,n\}.
\label{D.10}
\ee
Inspection of the signs of $w_a(\lambda)$ and $w_{a+n}(\lambda)$
in (\ref{d38}) gives
\be
w_a(\lambda)<0<w_{a+n}(\lambda).
\label{D.11}
\ee
Since every boundary point $\lambda^0\in\partial\overline{C}_2$ satisfies
$\lambda_a^0>\lambda_n^0\geq \nu$ for all $a\in\{1,\ldots,n-1\}$, we may choose
 a  small enough ball  centred at $\lambda^0$ to ensure that for
$\lambda$ inside that ball the above inequalities as well as
$\lambda_a>\nu$  hold.
On account of $\lambda_a>\nu>0$ and $\mu>0$  we then have
\be
1 - \frac{\nu}{\lambda_a}>0 \quad\hbox{and}\quad
-1-\frac{2\mu-\nu}{\lambda_a}<0.
\label{D.14}
\ee
By combining (4.41) and (4.42) with  (\ref{D.11}) and
\eqref{D.14} we conclude that
\be
\cF_a^+(\lambda)<0 \quad\hbox{and}\quad  \cF_{a+n}^-(\lambda)<0.
\label{bad1}
\ee
By Lemma 4.5,
these inequalities imply that
$\cF_a(\lambda)$ and $\cF_{a+n}(\lambda)$ cannot be both non-negative,
which contradicts the defining equation (\ref{d29}).
This proves the claim in the case (i) (\ref{D.6}).

Let us consider a  strongly regular $\lambda$  satisfying (ii)
\eqref{D.7}.
In this case we can verify that
\be
1-\frac{\nu}{\lambda_n}<0,\quad
w_n(\lambda)>0,\quad w_{n+a}(\lambda)>0.
\label{D.16}
\ee
Thus we see from (\ref{d41}) that $\cF_{2n}^+(\lambda)<0$.
Since the sum of the two components on the right hand side of (\ref{d42}) is negative, we also see
that at least one out of $\cF_{n}^-(\lambda)$ and $\cF_{2n}^-(\lambda)$ is negative.
Therefore equations (\ref{d39}) and (\ref{d40}) exclude the unitarity of $\check A$ (\ref{d37})
in the case (ii) (\ref{D.7}) as well.
\end{proof}
\medskip

\noindent\bf\sffamily Proposition 4.9. \it
The $\lambda$-image of the constraint surface is contained in
$\overline{C}_2$, i.e., we have
\be
\fL(P_0) \subseteq \overline{C}_2.
\label{im}
\ee
As a consequence, $\fL_\red^{-1}(C_2)$ is dense in $P_\red$.
\rm

\begin{proof}
Since $P_0^\vreg \subset P_0$ is dense and $\fL: P_0 \to \R^n$ (\ref{d8})
is continuous, $\fL(P_0^\vreg) \subset \fL(P_0)$ is dense.
Thus it follows from Lemma 4.8 that for any
$\lambda^0\in\partial C_2$ there exists
a ball around $\lambda^0$ that does not contain \emph{any} element of
$\fL(P_0)$ lying outside $\overline{C}_2$.

Suppose that (\ref{im}) is not true, which means that there exists some
$\lambda^*\in \fL(P_0)\setminus \overline{C}_2$. Taking any element
$\hat \lambda \in \fL(P_0)$ that lies in $C_2$, it is must be possible
to connect $\lambda^*$ to $\hat \lambda$
by a continuous curve in $\fL(P_0)$, since $P_0$ is connected.
Starting from the point $\lambda^*$, any such continuous curve
must pass through some point of the boundary $\partial C_2$.
However, this is impossible since we know that
$\fL(P_0)\setminus \overline{C}_2$ does not contain any series that
converges to a point of $\partial C_2$.
This contradiction shows that (\ref{im}) holds.

By (\ref{im}) we have $P_0^\vreg \subset \fL^{-1}(C_2)$, and we know from Lemma 4.7 that
$P_0^\vreg \subset P_0$ is dense. These together entail that $\fL_\red^{-1}(C_2)\subset P_\red $ is dense.
\end{proof}

\subsection{Global characterization of the dual system}

We have seen that
\be
P_0^\vreg \subset  \fL^{-1}(C_2) \subset P_0
\label{G4}
\ee
is a chain of dense open submanifolds.
These project onto dense open submanifolds of $P_\red$ and
their images under the  map $\fL$ (\ref{d8}) are
dense subsets of $\fL(P_0)=\fL_\red(P_\red)$:
\be
\fL(P_0^\vreg) \subset C_2 \subset \fL(P_0).
\label{G5}
\ee

Now introduce the set
\be
\C^n_{\neq }:= \{ z\in \C^n \mid \prod_{k=1}^n z_k \neq 0\}.
\label{G6}
\ee
The parametrization
\be
z_j=\sqrt{\lambda_j-\lambda_{j+1}-2\mu}\prod_{a=1}^j e^{\ri\vartheta_a},
\,\,\, j=1,\ldots, n-1, \qquad
z_n=\sqrt{\lambda_n-\nu}\prod_{a=1}^n e^{\ri\vartheta_a}
\label{G7}
\ee
provides a diffeomorphism between $C_2 \times \T^n$ and $\C^n_{\neq }$.
Thus we can view $z\in \C^n_{\neq}$ as a variable parametrizing
$C_2 \times \T^n$ that corresponds to the semi-global cross-section $\tilde S^0$
by Theorem 4.1. Below, we shall exhibit a \emph{global cross-section} in $P_0$,
which will be diffeomorphic to $\C^n$. In other words, the `semi-global'
model of the dual systems will be completed into a global model by allowing
the zero value for the complex variables $z_k$.  This completion results
from the symplectic reduction automatically.

First of all, let us note that the inverse of the parametrization (\ref{G7}) gives
\be
\lambda_k(z)=\nu+2(n-k)\mu+\sum_{j=k}^nz_j\bar z_j,\quad k=1,\ldots,n,
\label{G8}
\ee
which extend to smooth functions over $\C^n$.
The range of the extended map $z \mapsto (\lambda_1, \ldots, \lambda_n)$
is the closure $\overline{C}_2$ of the polyhedron $C_2$.
The variables $e^{\ri \vartheta_k}$ are well-defined only over
$\C^n_{\neq }$, where the parametrization (\ref{G7}) entails the equality
\be
\sum_{k=1}^n\rd \lambda_k \wedge\rd \vartheta_k
= \ri\sum_{k=1}^n\rd z_k \wedge\rd \bar z_k.
\label{G9}
\ee

An easy inspection of the formulae (\ref{d11}) shows that the
functions $f_a$ can be recast as
\be
f_k(\lambda, e^{\ri \vartheta}) = \vert z_k \vert g_k(z),
\quad
f_{n+k}(\lambda, e^{\ri \vartheta})
= e^{\ri \vartheta_k} \vert z_{k-1}\vert g_{n+k}(z),
\qquad k=1,\ldots, n,\,\,\, z_0:=1,
\label{G62}
\ee
with uniquely defined functions $g_1(z),\ldots, g_{2n}(z)$
that extend to smooth (actually real-analy\-tic) positive functions on $\C^n$.
Note that these functions depend on $z$ only through $\lambda(z)$, i.e.,
one has
\be
g_a(z) = \eta_a(\lambda(z)),
\qquad
a=1,\ldots, N,
\ee
with suitable functions $\eta_a$ that one could display explicitly.
The absolute values $\vert z_k \vert$ that appear in (\ref{G62}) are not
smooth at $z_k=0$, and the phases
$e^{\ri \vartheta_k}$ are not well-defined there.
The crux is that both of these `troublesome features' can be
removed by applying suitable gauge transformations to the elements
of the cross-section $\tilde S^0$ (\ref{d17}).
To demonstrate this, we define $m=  m(e^{\ri \vartheta})\in Z_{G_+}(\cA)$ by
\be
m_k(e^{\ri \vartheta}): = \prod_{j=1}^k e^{-\ri \vartheta_j},
\qquad
k=1,\ldots, n.
\label{G63}\ee
Conforming with (\ref{P28}), we also set $m_{k+n}=m_k$.
Then the  gauge transformation by $(m,m)\in G_+ \times G_+$ operates on the
$\C^N$-valued vector $f(\lambda, e^{\ri \vartheta})$ and on the matrix
$\check A(\lambda, e^{\ri \vartheta})$ according to
\be
f(\lambda, e^{\ri \vartheta}) \to
m(e^{\ri \vartheta}) f(\lambda, e^{\ri \vartheta}) \equiv \phi(z),
\qquad
\check A(\lambda, e^{\ri \vartheta)} \to m(e^{\ri \vartheta})
\check A(\lambda, e^{\ri \vartheta}) m(e^{\ri \vartheta})^{-1} \equiv \tilde A(z),
\ee
which defines the functions $\phi(z)$ and $\tilde A(z)$ over $\C^n_{\neq}$.
The resulting functions have the form
\be
\phi_k(z)= \bar z_k g_k(z), \quad
\phi_{n+k}(z) = \bar z_{k-1} g_{n+k}(z), \quad k=1,\ldots, n,
\ee
and
\be
\tilde{A}_{a,b}(z)
=-\frac{2\mu\bar{z}_az_{b-1}g_a(z)g_{n+b}(z)}
{\lambda_a(z)-\lambda_b(z)-2\mu}, \quad 1\leq a,b \leq n,
\label{G66}
\ee
\be
\tilde{A}_{a,n+b}(z)
=-\frac{2\mu\bar{z}_az_bg_a(z)g_b(z)}
{\lambda_a(z)+\lambda_b(z)-2\mu}
+\delta_{a,b}\frac{\mu -\nu}{\lambda_a(z)-\mu},
\ee
\be
\tilde{A}_{n+a,b}(z)
=\frac{2\mu\bar{z}_{a-1}z_{b-1}g_{n+a}(z)g_{n+b}(z)}
{\lambda_a(z)+\lambda_b(z)+2\mu}
-\delta_{a,b}\frac{\mu-\nu}{\lambda_a(z)+\mu},
\ee
\be
\tilde{A}_{n+a,n+b}(z)
=\frac{2\mu\bar{z}_{a-1}z_b g_{n+a}(z)g_b(z)}
{\lambda_a(z)-\lambda_b(z)+2\mu}.
\ee
Now the important point is that, as is easily verified, the apparent
singularities coming from vanishing denominators in $\tilde A$  all cancel,
and both $\phi(z)$ and $\tilde A(z)$ extend to smooth (actually real-analytic) functions on
the whole of $\C^n$.
In particular, note the relation
\be
\tilde{A}_{k,k+1}(z)= \tilde A_{k+n+1,k+n}(z) = - 2\mu g_k(z) g_{k+n+1}(z),
\qquad
k=1,\ldots, n-1.
\label{G71new}
\ee
Corresponding to (\ref{d13}), we also have the matrix $\tilde B(z)\equiv-(h(\lambda(z))
\tilde A(z)h(\lambda(z)))^\dagger$.
This is smooth over $\C^n$ since both $\tilde A(z)$ and $h(\lambda(z))$ (\ref{d5}) are smooth.
It follows from their defining equations that the induced gauge transformations
of $y(\lambda, e^{\ri \vartheta})$ (\ref{d15}) and $V(\lambda, e^{\ri \vartheta})$ (\ref{d16})
are given by
\be
y(\lambda, e^{\ri \vartheta}) \to m(e^{\ri \vartheta})
y(\lambda, e^{\ri \vartheta}) m(e^{\ri \vartheta})^{-1} \equiv \tilde y(z),
\ee
and
\be
V(\lambda, e^{\ri \vartheta}) \to m(e^{\ri \vartheta}) V(\lambda, e^{\ri \vartheta})
= \tilde y(z) h(\lambda(z)) \phi(z)
\equiv \tilde V(z).
\ee
Since $\tilde y(z)$ is a uniquely defined smooth function of $\tilde B(z)$, both $\tilde y(z)$ and
$\tilde V(z)$ are smooth functions on the whole of $\C^n$.

After these preparations, we are ready to state the main result of this paper.

\medskip
\noindent \bf\sffamily Theorem 4.10. \it
By using the above notations, consider the set
\be
\tilde S:= \{( \tilde y(z), \ri h(\lambda(z)) \Lambda(\lambda(z)) h(\lambda(z))^{-1},
\upsilon^\ell_{\mu, \nu}(\tilde V(z)),
\upsilon^r)\mid z\in
\C^n\,\}.
\label{G72}
\ee
This set defines a global cross-section for the $G_+\times G_+$-action
on the constraint surface $P_0$.
The parametrization of the elements of $\tilde S$  by $z\in \C^n$ gives
rise to a symplectic diffeomorphism between $(P_\red, \Omega_\red)$ and
$\C^n$ equipped with the Darboux form $\ri \sum_{k=1}^n\rd z_k\wedge\rd \bar z_k$.
The spectral invariants of the  `global RSvD Lax matrix'
\be
\tilde L(z) \equiv h(\lambda(z)) \tilde A(z) h(\lambda(z))
\label{G73}
\ee
yield commuting Hamiltonians on $\C^n$ that represent the reductions of
the Hamiltonians spanning the Abelian Poisson algebra $\fQ^1$ (\ref{P23}).
\rm

\begin{proof}
Let us denote by
\be
z \mapsto \tilde \sigma(z)
\label{G74}
\ee
the assignment of the element of $\tilde S$ to $z\in \C^n$ as given in (\ref{G72}).
The map $\tilde\sigma\colon\C^n\to P$ (\ref{P17}) is smooth (even real-analytic)
and we have to verify that it possesses the following properties.
First, $\tilde\sigma$ takes values in the constraint surface $P_0$.
Second, with $\Omega$ in (\ref{P17}),
\be
\tilde \sigma^*(\Omega) = \ri \sum_{k=1}^n\rd z_k\wedge\rd \bar z_k.
\label{G75}
\ee
Third, $\tilde\sigma$ is injective.
Fourth, the image $\tilde S$ of $\tilde\sigma$ intersects every orbit of $G_+ \times G_+$
in $P_0$ in precisely one point.

Let us start by recalling from Theorem 4.1 the map
$(\lambda, \theta)\mapsto \tilde \sigma_0(\lambda, \theta)$
that denotes the assignment of the general element of $\tilde S^0$ (\ref{d17}) to
$(\lambda, \theta) \in C_2 \times \T^n$, where now we defined
\be
\theta:= e^{\ri \vartheta}.
\label{G77}
\ee
Then the first and second  properties of $\tilde \sigma$ follow since we have
\be
\tilde \sigma(z(\lambda, \theta)) = \Phi_{(m(\theta), m(\theta))}
\left(\tilde \sigma_0(\lambda, \theta)\right),
\quad
\text{for all}\quad (\lambda, \theta) \in C_2 \times \T^n.
\label{G78}
\ee
We know that $\tilde \sigma_0(\lambda, \theta) \in P_0$ for all $(\lambda, \theta)\in
C_2 \times \T^n$, which implies the first property since $\tilde\sigma$ is continuous and
$P_0$ is a closed subset of $P$. The restriction of the pull-back (\ref{G75})
to $\C^n_{\neq}$ is easily calculated using the parametrization $(\lambda, \theta) \mapsto
z(\lambda, \theta)$ and using that by Theorem 4.1
$\tilde \sigma_0^*(\Omega)=\sum_{k=1}^n\rd\lambda_k \wedge\rd\vartheta_k$.
Indeed, this translates into (\ref{G75}) restricted to $\C^n_{\neq}$,
which implies the claimed
equality because $\tilde \sigma^*(\Omega)$ is smooth on $\C^n$.

Before continuing, we remark that the map $(\lambda, \theta) \mapsto z(\lambda, \theta)$
naturally extends to a continuous map on the closed domain $\overline{C}_2 \times \T^n$ and its
`partial inverse' $z\mapsto\lambda(z)$ extends to a smooth map $\C^n\to\overline{C}_2$.
We will use these extended maps without further notice in what follows.
(The extended map $(\lambda, \theta) \mapsto z(\lambda, \theta)$ is not
differentiable at the points for which $\lambda\in\partial C_2$.)

In order to show that $\tilde \sigma$ is injective, consider the equality
\be
\tilde \sigma(z) = \tilde \sigma(\zeta)\quad
\hbox{for some}\quad z, \zeta \in \C^n.
\ee
Looking at the `second component' of this equality according to (\ref{G72}) we see
that $\lambda(z) = \lambda(\zeta)$. Then the first component of the equality implies
$\tilde A(z) = \tilde A(\zeta)$. The special case $\tilde A_{a,1}(z) = \tilde A_{a,1}(\zeta)$
of this equality gives
\be
\frac{\bar{z}_a \eta_a(\lambda(z))\eta_{n+1}(\lambda(z))}
{\lambda_a(z)-\lambda_1(z)-2\mu}
=\frac{\bar{\zeta}_a \eta_a(\lambda(\zeta))\eta_{n+1}(\lambda(\zeta))}
{\lambda_a(\zeta)-\lambda_1(\zeta)-2\mu}, \qquad 1\leq a \leq n.
\ee
We know that the factors multiplying $\bar z_a$ and $\bar \zeta_a$ are
equal and non-zero (actually negative). Thus $z=\zeta$ follows,
establishing the claimed injectivity.

Next we prove that no two different element of $\tilde S$ are gauge equivalent
to each other, i.e., $\tilde S$ can intersect any orbit of $G_+ \times G_+$
at most in one point.
Suppose that
\be
\Phi_{(g_L, g_R)}( \tilde \sigma(z)) = \tilde \sigma(\zeta)
\label{G82}
\ee
for some $(g_L, g_R) \in G_+ \times G_+$ and $z, \zeta \in \C^n$.
We conclude from the second component of this equality that $\lambda(z)= \lambda(\zeta)$.
Because $\lambda(z)\in\overline{C}_2$ holds, $\lambda(z)$ is
regular in the sense that it satisfies (\ref{d23}).  Thus we can also conclude
from the second component of the equality (\ref{G82})  that $g_R$ belongs to
the Abelian subgroup $Z$ of $G_+$ given in (\ref{P28}).
Then we infer from the first component
\be
g_L \tilde y(z) g_R^{-1} =  \tilde y(\zeta)
\ee
of the equality (\ref{G82})  that $g_L = g_R$.
We here used that $\tilde A(\zeta)$ can be represented in the form (\ref{P32}) with
strict inequalities in (\ref{P30}), which holds since $S$ (\ref{S6}) is a global cross-section.
Now denote $ g_L = g_R = e^{\ri \xi} \in Z$ referring to (\ref{P28}).
Then we have $e^{\ri \xi} \tilde A(z) e^{-\ri \xi} = \tilde A(\zeta)$,
and in particular
\be
e^{\ri x_a} \tilde A_{a, a+1}(z)  e^{-\ri x_{a+1}} = \tilde A_{a,a+1}(\zeta),
\quad
\forall a=1,\ldots, n-1.
\ee
By using (\ref{G62}) and (\ref{G71new})
\be
\tilde{A}_{a,a+1}(z)
=-2\mu \eta_a(\lambda(z))\eta_{n+a+1}(\lambda(z)) \neq 0,
\ee
and thus we obtain from $\lambda(z) = \lambda(\zeta)$ that $ e^{\ri \xi}$ must be equal to a multiple
of the identity element of $G_+$.
Hence we have established that $\tilde \sigma(z) = \tilde \sigma (\zeta)$ is implied by
(\ref{G82}).

It remains to demonstrate that $\tilde S$ intersects every gauge orbit in $P_0$.
We have seen previously that $\fL^{-1}(C_2)$ is dense in $P_0$
and $\tilde S^0$ (\ref{d17}) is a cross-section for the gauge action in $\fL^{-1}(C_2)$.
These facts imply that for any element $x \in P_0$ there exists a series
$x(k) \in \fL^{-1}(C_2)$, $k\in \N$,  such that
\be
\lim_{k\to \infty}(x(k)) =x,
\ee
and there also exist series $(g_L(k), g_R(k))\in G_+ \times G_+$ and $(\lambda(k),\theta(k))
\in C_2 \times \T^n$ such that
\be
x(k) = \Phi_{(g_L(k), g_R(k))}\left(\tilde \sigma_0( \lambda(k), \theta(k))\right).
\ee
Since $\fL: P_0 \to  \R^n$ is continuous, we have
\be
\fL(x) = \lim_{k\to \infty} \fL(x(k)) = \lim_{k\to \infty} \lambda(k).
\ee
This limit belongs to $\overline{C}_2$ and we denote it by $\lambda^\infty$.
The non-trivial case to consider is when $\lambda^\infty$ belongs to the boundary
$\partial C_2$.
Now, since $G_+ \times G_+ \times \T^n$ is compact, there exists a
convergent subseries
\be
(g_L(k_i), g_R(k_i), \theta(k_i)),
\quad
i \in \N,
\ee
of the series $(g_L(k), g_R(k), \theta(k))$.
We pick such a convergent subseries and denote its limit as
\be
(g_L^\infty, g_R^\infty, \theta^\infty):=\lim_{i\to \infty} (g_L(k_i), g_R(k_i), \theta(k_i)).
\ee
Then we define $z^\infty \in \C^n$ by
\be
z^\infty := \lim_{i\to \infty} z(\lambda(k_i), \theta(k_i)) = z(\lambda^\infty, \theta^\infty).
\ee
Since $z \mapsto \tilde \sigma(z)$ is continuous, we can write
\be
\tilde \sigma(z^\infty) = \lim_{i\to \infty} \tilde \sigma( z(\lambda(k_i), \theta(k_i)))=
\lim_{i\to \infty} \Phi_{(m(\theta(k_i)), m(\theta(k_i)))}
\left(\tilde \sigma_0(\lambda(k_i), \theta(k_i))\right),
\ee
where $m(\theta)$ is defined by (\ref{G63}), with $\theta = e^{\ri \vartheta}$.
By combining these formulae, we finally obtain
\bea
&& x = \lim_{i\to \infty} \Phi_{(g_L(k_i),g_R(k_i))}
\left(\tilde \sigma_0(\lambda(k_i), \theta(k_i))\right)
\nonumber\\
&& \quad = \lim_{i\to \infty} \Phi_{(g_L(k_i) m(\theta(k_i))^{-1},g_R(k_i)m(\theta(k_i))^{-1}  )}
\left(\tilde \sigma( z(\lambda(k_i), \theta(k_i)))\right)\nonumber \\
&&\quad
= \Phi_{( g_L^\infty m(\theta^\infty)^{-1}, g_R^\infty m(\theta^\infty)^{-1})}
(\tilde \sigma (z^\infty)).
\eea
Therefore $\tilde S$ is a global cross-section in $P_0$.

The final statement of Theorem 4.10 about the global RSvD Lax matrix (\ref{G73}) follows since
$\tilde L$ is just the restriction of the `unreduced Lax matrix'  $L$ of (\ref{d2})
to the global cross-section $\tilde S$, which represents a model of the full reduced phase space
$P_\red$.
\end{proof}

\section{Discussion}
\setcounter{equation}{0}

In this paper we characterized a symplectic reduction of the phase space
$(P, \Omega)$ (\ref{P17}) by exhibiting two models of the reduced phase space
$P_\red$ (\ref{P20}).  These are provided by the global cross-sections $S$ and
$\tilde S$
described in Theorem 3.1 and in Theorem 4.10. The two cross-sections naturally give rise
to symplectomorphisms
\be
(M, \omega) \simeq (P_\red, \Omega_\red) \simeq (\tilde M, \tilde \omega),
\label{F1}
\ee
where $M= T^* C_1$ (\ref{I2}) with the canonical
symplectic form $\omega= \sum_{k=1}^n\rd q_k \wedge\rd p_k$
and $\tilde M = \C^n$ with $\tilde \omega=\ri \sum_{k=1}^n\rd z_k \wedge\rd\bar z_k$.
The Abelian Poisson algebras $\fQ^1$ and $\fQ^2$ on $P$ (\ref{P23}) descend to reduced
Abelian Poisson algebras
$\fQ^1_\red$ and $\fQ^2_\red$ on $P_\red$.
The construction guarantees that
any element of the reduced Abelian Poisson algebras possesses complete Hamiltonian flow.
These flows  can be analyzed by means of the standard projection
algorithm as well as by utilization of the symplectomorphism (\ref{F1}).

To further discuss the interpretation of our results,
consider the gauge invariant functions
\be
\cH_k(y,Y,V) = \frac{1}{4k} (-\ri Y)^{2k}
\quad\text{and}\quad
{\tilde \cH}_k(y,Y,V) = \frac{(-1)^k}{2k} \tr(y^{-1} C y C)^k,
\quad
k=1,\ldots, n.
\label{F2}
\ee
The restrictions of the functions $\cH_k$ to the global
cross-sections $S$ and $\tilde S$ take the form
\be
\cH_k|_{S}=\frac{1}{4k} (-\ri Y(q,p))^{2k} = H_k(q,p)
\quad\text{and}\quad
\cH_k|_{\tilde S}=\frac{1}{2k} \sum_{j=1}^n \lambda_j(z)^{2k}.
\label{F3}
\ee
According to  (\ref{S19}), the $H_k$ yield the commuting Hamiltonians of the
Sutherland system,
while the $\lambda_j$ as functions on $\tilde S\simeq \C^n$ are given by (\ref{G8}).
Since any smooth function on a global cross-section encodes a smooth function on
$P_\red$, we conclude
that the Sutherland Hamiltonians $H_k$ and the `eigenvalue-functions'
$\lambda_j$ define two alternative sets of generators for $\fQ^2_\red$.

The restrictions of the functions $\tilde \cH_k$ read
\be
{\tilde \cH}_k \vert_{S} = \frac{(-1)^k}{k}\sum_{j=1}^n \cos( 2k q_j)
\quad\hbox{and}\quad
{\tilde \cH}_k\vert_{\tilde S}=\frac{1}{2k}\tr(\tilde L(z)^k)
\label{F4}
\ee
with $\tilde L(z)$ is defined in (\ref{G73}).
On the semi-global cross-section $\tilde S^0$ of Theorem 4.1,
which parametrizes the dense open submanifold
$\fL^{-1}_\red(C_2) \subset P_\red$,
we have
\be
{\tilde \cH}_1\vert_{\tilde S^0}= \tilde H^0,
\label{F5}
\ee
where $\tilde H^0$ is the RSvD Hamiltonian displayed in (\ref{I4}).
We see from (\ref{F4}) that
the functions $q_j\in C^\infty(S)$ and the commuting
Hamiltonians ${\tilde \cH}_k|_{\tilde S}$ engender two alternative generating sets for
$\fQ^1_\red$.
On account of the relations
\be
\tilde M^0\simeq\tilde S^0\simeq C_2\times\T^n\simeq\C^n_{\neq}\subset\C^n\simeq\tilde S\simeq\tilde M,
\label{F6}
\ee
${\tilde \cH}_1\vert_{ \tilde S}$ yields a globally smooth extension of the
many-body Hamiltonian $\tilde H^0$.

It is immediate from our results that both $\fQ^1_\red$ and $\fQ^2_\red$
define Liouville integrable systems on $P_\red$, since both
have $n$ functionally independent generators.
The interpretations of these Abelian Poisson algebras that stem from the models
$S$ and $\tilde S$ underlie the action-angle duality between the Sutherland and RSvD systems
as follows. First,
the generators $q_k$ of $\fQ^1_\red$
can be viewed alternatively as particle positions for the Sutherland system or as
action variables for the RSvD system.
Their canonical conjugates $p_k$ are of non-compact type.
Second, the generators $\lambda_k$ of $\fQ^2_\red$ can be viewed alternatively as action variables
for the Sutherland systems or as globally well-defined `particle positions' for the
completed RSvD system.
In conclusion, the symplectomorphism $\cR: M \to \tilde M$ naturally
induced by (\ref{F1}) satisfies all properties required by the notion of action-angle
duality outlined in the Introduction.

We finish by pointing out some further consequences.
First of all, we note that the dimension
of the Liouville tori of the Sutherland system drops on the locus where
the action variables encoded by $\lambda$  belong to the boundary of the
polyhedron $\overline{C}_2$. This is a consequence of the next statement, which can be proved
by direct calculation.

\medskip
\noindent \bf\sffamily Proposition 5.1. \it
Consider the Sutherland Hamiltonians\footnote{Here
$H_k(z)$ denotes the reduction
of the Hamiltonian $\cH_k$ expressed in terms of the model $\tilde M$, cf.~(\ref{F3}).}
$H_k(z) = \frac{1}{2k} \sum_{j=1}^n \lambda_j(z)^{2k}$
and for any $z \in \C^n$ define $\cD(z):= \# \{ z_k\neq 0\mid k=1,\ldots, n\}$.
Then one has the equality
\be
\operatorname{dim}\left( \operatorname{span} \{\rd\lambda_k(z) \mid k=1,\ldots,n \}\right)
=\operatorname{dim}\left( \operatorname{span}
 \{\rd H_k(z)  \mid k=1,\ldots,n \}\right)= \cD(z).
\label{F7}
\ee
\rm

It follows from (\ref{F7}) that the dense open submanifold
$\fL_\red^{-1}(C_2) \subset P_\red$ corresponds to the part of
the Sutherland phase space where the Liouville tori have full dimension $n$.
It is also worth noting that the special point for which $z=0$, or equivalently
\be
\lambda_j = \nu + 2 (n-j) \mu,
\quad
\forall j=1,\ldots, n,
\label{F8}
\ee
gives the unique global minimum of the function $H_1(z)$.
Equation (\ref{F3}) implies that actually each function $H_k$ ($k=1,\ldots, n$)
possesses a global minimum at $z=0$.
An interesting characterization of this equilibrium point in terms of the $(q,p)$ variables can
be found in \cite{CS}.

Being in control of the action-angle variables for our dual pair of integrable
systems, the following result is readily obtained.

\medskip
\noindent \bf\sffamily Proposition 5.2. \it
Any `Sutherland Hamiltonian' $H_k \in C^\infty(M)$  ($k=1,\ldots, n$)
given by (\ref{S18})
defines a non-degenerate Liouville integrable system, i.e.,
the commutant of $H_k$ in the Poisson algebra $C^\infty(M)$ is the Abelian
algebra generated by the action variables $\lambda_1,\ldots, \lambda_n$.
Any `RSvD Hamiltonian' $\tilde H_k \in C^\infty(\tilde M)$, $k=1,\ldots, n$,
 which by definition coincides with $\tilde \cH_k\vert_{\tilde S}$ in (\ref{F4}) upon the
identification $\tilde M  = \tilde S$,
is maximally degenerate (`superintegrable') since its commutant in the
Poisson algebra $C^\infty(\tilde M)$ is generated by $(2n-1)$ elements.
\rm
\begin{proof}
The subsequent argument relies on the `action-angle symplectomorphisms'
 between $(M,\omega)$ and $(\tilde M, \tilde \omega)$  corresponding to (\ref{F1}).

Let us  first restrict the Sutherland Hamiltonian $H_k$
to the submanifold parametrized by the action-angle variables
varying in $C_2 \times \T^n$.
For generic $\lambda$, we see from \eqref{F3} that the flow of $H_k$ is dense on the torus $\T^n$.
Therefore any smooth function $f$ that Poisson commutes with $H_k$ must
be constant on the non-degenerate Liouville tori of the Sutherland system.
By smoothness, this implies that $f$ Poisson commutes with all the action variables $\lambda_j$ on
the full phase space.
Consequently, it can be expressed as a function of those variables.

Next, by a slight abuse of notation, let us  write $\tilde H_k(q,p)= \tilde h_k(q)$ for  the `RSvD Hamiltonian' expressed
in terms of the associated `dual action-angle phase space' $M= C_1 \times \R^n$.
By   (\ref{F4}),
$\tilde h_k(q) = \frac{(-1)^k}{k} \sum_{j=1}^n \cos(2k q_j)$
and one can  verify that the matrix
\be
X_{i,j}(q):= \frac{\partial \tilde h_i(q)}{\partial q_j}
\ee
is non-degenerate for all $q\in C_1$.
As argued in \cite{AFG}, this implies that $\tilde H_k$
is maximally superintegrable. In fact, the commutant of $\tilde H_k$  is generated by
the `dual actions' $q_1, \ldots, q_n$ together with the functions
\be
f_i(q,p):= \sum_{j=1}^n p_j (X(q))^{-1}_{j,i},\quad
i\in \N_n\setminus \{k\}.
\ee
This concludes the proof.
\end{proof}

In the end,
we remark that the matrix functions $-\ri Y(q,p)$ and $\tilde L(z)$,
which naturally arose from the Hamiltonian reduction,
serve as Lax matrices for the pertinent dual pair of integrable systems.
We also notice that the $z_j$  can be viewed as
 `oscillator variables'
for the Sutherland system since the actions $\lambda_k$ are linear combinations
 in $\vert z_j \vert^2$ ($j=1,\ldots, n$) and
the form $\tilde\omega$ coincides with the symplectic form of $n$ independent harmonic
oscillators.
It could be worthwhile to inspect the quantization of
 the Sutherland system based on these oscillator variables
and to compare the result to the standard quantization \cite{H,HO,O}.
We plan to return to this issue in the future.

\bigskip
\bigskip
\bigskip
\noindent{\bf Acknowledgements.}
This work was supported in part by the Hungarian Scientific Research
Fund (OTKA) under the grant
K-111697 and by the project  T\'AMOP-4.2.2.A-11/1/KONV-2012-0060
financed by the EU and co-financed by the European Social Fund.
TFG's research was also supported by the EU and the State of Hungary,
co-financed by the European Social Fund in the framework of T\'AMOP-4.2.4.A/
2-11/1-2012-0001 `National Excellence Program'.

\bigskip

\renewcommand{\theequation}{\arabic{section}.\arabic{equation}}
\renewcommand{\thesection}{\Alph{section}}
\setcounter{section}{0}

\section{Some technical details}
\setcounter{equation}{0}
\renewcommand{\theequation}{A.\arabic{equation}}

In this appendix we complete the proof of Lemma 4.5 by
a calculation based on Jacobi's theorem on
complementary minors (e.g.~\cite{Pras}), which  will be recalled shortly.
Our reasoning below is adapted from Pusztai \cite{P1}.
A significant difference is that in our case we need the strong regularity conditions
(\ref{d23}) and (\ref{d36}) to avoid dividing by zero during
the calculation. In fact, this appendix is presented mainly
to explain the origin of the strong regularity conditions.

For an $m\times m$ matrix $M$ let
$M\big(\begin{smallmatrix}r_1&\cdots&r_k\\c_1&\cdots&c_k\end{smallmatrix}\big)$
denote the determinant formed from the entries lying on the intersection of
the rows $r_1,\ldots,r_k$ with the columns $c_1,\ldots,c_k$ of $M$ $(k\leq m)$,
$$M\begin{pmatrix}r_1&\cdots&r_k\\c_1&\cdots&c_k\end{pmatrix}
=\det(M_{r_i,c_j})_{i,j=1}^k.$$

\medskip
\noindent\bf\sffamily Theorem A.1 (Jacobi). \it
Let $A$ be an invertible $N\times N$ matrix with $\det(A)=1$ and $B:=(A^{-1})^\top$.
For a fixed permutation
$\big(\begin{smallmatrix}j_1&\cdots&j_N\\k_1&\cdots&k_N\end{smallmatrix}\big)$
of the pairwise distinct indices $j_1,\ldots,j_N\in\{1,\ldots,N\}$ and
any $1\leq p<N$
\be
B\begin{pmatrix}j_1&\cdots&j_p\\k_1&\cdots&k_p\end{pmatrix}
=\sgn\begin{pmatrix}j_1&\cdots&j_N\\k_1&\cdots&k_N\end{pmatrix}
A\begin{pmatrix}j_{p+1}&\cdots&j_N\\k_{p+1}&\cdots&k_N\end{pmatrix}.
\label{A.1}
\ee
\rm

Applying Jacobi's theorem to $\check A$ (4.37) we now derive
the two equations (\ref{d39}) and (\ref{d40})
for the pair of functions $(W_a,W_{n+a})$ for each $a=1,\ldots, n$, which are defined
by $W_k = w_k \cF_k$ with $\cF_k = \vert F_k \vert^2$ \eqref{d29} and $w_k$ in (\ref{d38}).

\noindent\bf\sffamily Lemma A.2. \it
Fix any  strongly regular $\lambda$, i.e., $\lambda\in\R^n$
for which (\ref{d23}) and (\ref{d36}) hold,
and use the above notations for $(W_a,W_{n+a})$.
If $\check A$ given by (\ref{d37}) is a unitary matrix, then
$(W_a,W_{n+a})$ satisfies the two equations (\ref{d39}) and
(\ref{d40}) for each $a=1,\ldots, n$.
\rm
\begin{proof}
Let $\check B:=(\check{A}^{-1})^\top$, i.e. $\check{B}_{j,k}:=\overline{\check{A}}_{j,k}$, $j,k\in\{1,\ldots,N\}$
and $a\in\{1,\ldots,n\}$ be a fixed index. Since $\det(\check A)=1$, by
Jacobi's theorem with $j_b=b$, ($b\in\N_N$) and $k_c=c$, ($c\in\N_N\setminus\{a,n+a\}$), $k_a=n+a$, $k_{n+a}=a$ and $p=n$
we have
\be
\check{B}\begin{pmatrix}
1&\cdots&a&\cdots&n\\
1&\cdots&n+a&\cdots&n
\end{pmatrix}=
-\check{A}\begin{pmatrix}
n+1&\cdots&n+a&\cdots&N\\
n+1&\cdots&a&\cdots&N
\end{pmatrix}.
\label{A.3}
\ee
Denote the corresponding $n\times n$  submatrices of $\check B$ and $\check A$
 by $\xi$ and $\eta$, respectively.
One can check that
\be
\xi=\Psi-\frac{\mu-\nu}{\mu-\lambda_a}E_{a,a},\quad
\eta=\Xi-\frac{\mu-\nu}{\mu+\lambda_a}E_{a,a},
\label{A.4}
\ee
where $E_{j,k}$ stands for the $n\times n$ elementary matrix
$(E_{j,k})_{j',k'}=\delta_{j,j'}\delta_{k,k'}$
and $\Psi$ and $\Xi$ are the Cauchy-like matrices
\be
\Psi_{j,k}:=\begin{cases}
\dfrac{2\mu\overline{F}_jF_{n+k}}{2\mu-\lambda_j+\lambda_k},&\text{if}\ k\neq a,\\[.5cm]
\dfrac{2\mu\overline{F}_jF_a}{2\mu-\lambda_j-\lambda_a},&\text{if}\ k=a,
\end{cases}
\qquad\text{and}\qquad
\Xi_{j,k}:=\begin{cases}
\dfrac{2\mu F_{n+j}\overline{F}_k}{2\mu+\lambda_j-\lambda_k},&\mbox{if}\ k\neq a,\\[.5cm]
\dfrac{2\mu F_{n+j}\overline{F}_{n+a}}{2\mu+\lambda_j+\lambda_a},&\mbox{if}\ k=a,
\end{cases}
\label{A.5}
\ee
$j,k\in\{1,\ldots,n\}$.
Expanding $\det(\xi)$ and $\det(\eta)$ along the $a$-th column we obtain the formulae
\be
\det(\xi)=\det(\Psi)-\frac{\mu-\nu}{\mu-\lambda_a}\cC_{a,a},\quad
\det(\eta)=\det(\Xi)-\frac{\mu-\nu}{\mu+\lambda_a}\cC_{a,a},
\label{A.6}
\ee
where $\cC_{a,a}$ is the cofactor of $\Psi$ associated with entry $\Psi_{a,a}$.
Since $\Psi$ and $\Xi$ are both Cauchy-like matrices we have
\be
\det(\Psi)=\frac{1}{\mu-\lambda_a}D_aW_a,\quad
\det(\Xi) =\frac{1}{\mu+\lambda_a}D_aW_{n+a},
\label{A.7}
\ee
where
\be
D_a:=\prod_{\substack{b=1\\(b\neq a)}}^n\overline{F}_bF_{n+b}
\prod_{\substack{c,d=1\\(a\neq c\neq d\neq a)}}^n
\frac{\lambda_c-\lambda_d}{2\mu+\lambda_c-\lambda_d}.
\label{A.8}
\ee
It can be easily seen that $\cC_{a,a}=D_a$, therefore formulae
\eqref{A.3}, \eqref{A.6}, \eqref{A.7} lead to the equation
\be
(\mu+\lambda_a)W_a+(\mu-\lambda_a)W_{n+a}-2(\mu-\nu)=0.
\label{A.9}
\ee

It should be noticed that in the last step we divided by $D_a$,
which is legitimate since $D_a$ is non-vanishing due to the strong-regularity
condition given by (\ref{d23}) and (\ref{d36}). To see this, assume momentarily that
$F_i=0$ for some $i=1,\ldots, n$ at some strongly regular $\lambda$.
The denominator in (\ref{d37}) does not vanish, and the unitarity of
$\check A$ implies that we must have $\check A_{i, i+n}=1$ or $\check A_{i, i+n}=-1$.
These in turn are equivalent to
\be
\lambda_i = 2\mu -\nu
\quad\hbox{or}\quad \lambda_i =\nu,
\ee
which are excluded by (\ref{d36}). One can similarly check that the vanishing of
$F_{n+i}$ would require
\be
\lambda_i =\nu - 2\mu
\quad\hbox{or}\quad \lambda_i =-\nu,
\ee
which are also excluded. These remarks pinpoint the origin of the second half of
the conditions imposed in (\ref{d36}).

Next, we apply Jacobi's theorem  by setting
$j_b=k_b=b$, ($b\in\N_n$), $j_{n+1}=k_{n+1}=n+a$, $j_{n+c}=k_{n+c}=n+c-1$,
($c\in\N_{n-1}$) and $p=n+1$. Thus
\be
\check{B}\begin{pmatrix}
1&\cdots&n&n+a\\
1&\cdots&n&n+a
\end{pmatrix}=
\check{A}\begin{pmatrix}
n+1&\cdots&\widehat{n+a}&\cdots&N\\
n+1&\cdots&\widehat{n+a}&\cdots&N
\end{pmatrix},
\label{A.10}
\ee
where $\widehat{n+a}$ indicates that the $(n+a)$-th row and column are omitted.
Now denote the submatrices of size $(n+1)$ and $(n-1)$ corresponding to the determinants in \eqref{A.10}
 by $X$ and $Y$, respectively.
From \eqref{A.10} and (\ref{d37}) it follows that $\det(X)=\det(Y)=D_a$ \eqref{A.8}.
The submatrix $X$ can be written in the form
\be
X=\Phi-\frac{\mu-\nu}{\mu-\lambda_a}E_{a,n+1}-\frac{\mu-\nu}{\mu+\lambda_a}E_{n+1,a},
\label{A.11}
\ee
i.e., $X$ is a rank two perturbation of the Cauchy-like matrix $\Phi$
having the entries
\be\begin{gathered}
\Phi_{j,k}:=\frac{2\mu\overline{F}_jF_{n+k}}{2\mu-\lambda_j+\lambda_k},\quad
\Phi_{j,n+1}:=\frac{2\mu\overline{F}_jF_a}{2\mu-\lambda_j-\lambda_a},\\
\Phi_{n+1,k}:=\frac{2\mu\overline{F}_{n+a}F_{n+k}}{2\mu+\lambda_a+\lambda_k},\quad
\Phi_{n+1,n+1}:=\overline{F}_{n+a}F_{a},
\label{A.12}
\end{gathered}\ee
where $j,k\in\{1,\ldots,n\}$.
The determinant of $\Phi$ is
\be
\det(\Phi)=-\frac{\lambda_a^2}{\mu^2-\lambda_a^2}D_a W_a W_{n+a},
\label{A.14}
\ee
which cannot vanish because $\lambda$ is strongly regular.
Since $X$ is a rank two perturbation of $\Phi$ we obtain
\be
\det(X)=\det(\Phi)-(\mu-\nu)\bigg(\frac{\cC_{a,n+1}}{\mu-\lambda_a}+\frac{\cC_{n+1,a}}{\mu+\lambda_a}\bigg)
+(\mu-\nu)^2\frac{\cC_{a,n+1}\cC_{n+1,a}-\cC_{a,a}\cC_{n+1,n+1}}{(\mu-\lambda_a)(\mu+\lambda_a)\det(\Phi)},
\label{A.13}
\ee
where $\cC$ now is used to denote the cofactors of $\Phi$.
By calculating the necessary cofactors we derive
\be
\begin{gathered}
\cC_{a,a}\cC_{n+1,n+1}=D_a^2W_aW_{n+a},\\
\cC_{a,n+1}=-\frac{1}{\mu+\lambda_a}D_aW_{n+a},\quad
\cC_{n+1,a}=-\frac{1}{\mu-\lambda_a}D_aW_a.
\end{gathered}
\label{A.15}
\ee
Equations \eqref{A.14}--\eqref{A.15} together with $\det(X)=D_a$ imply
\be
\lambda_a^2(W_aW_{n+a}-1)-\mu(\mu-\nu)(W_a+W_{n+a}-2)+\nu^2=0.
\label{A.16}
\ee
Equations (\ref{A.9}) and (\ref{A.16}) coincide with (\ref{d39}) and (\ref{d40}),
respectively.
\end{proof}

\end{document}